\documentclass[prb,preprint,showpacs,superscriptaddress,groupedaddress]{revtex4}  
\usepackage{makeidx}

\usepackage[colorlinks,hyperindex]{hyperref}
\usepackage[absolute]{textpos}
\usepackage{graphicx}  
\usepackage{epstopdf}
\usepackage{dcolumn}   
\usepackage{bm}        
\usepackage{amssymb}   
\hypersetup
{
	colorlinks,%
	citecolor=blue,%
	linkcolor=blue,%
	urlcolor=blue,%
}

\makeindex

\begin{document}
	
	
	\begin{textblock}{14}(7,1)
		\noindent Published: \href{http://journals.aps.org/prb/abstract/10.1103/PhysRevB.93.045202}{Physical Review B 93, 045202 (2016)}
	\end{textblock}

\title{Quantum mechanical prediction of four-phonon scattering rates and reduced thermal conductivity of solids}

\author{Tianli Feng}
\author{Xiulin Ruan}
\email{ruan@purdue.edu}
\affiliation{School of Mechanical Engineering and the Birck Nanotechnology Center, Purdue University, West Lafayette, Indiana 47907-2088, USA}
\date{\today}


\begin{abstract} 
	Recently, first principle-based predictions of lattice thermal conductivity $\kappa$ from perturbation theory have achieved significant success. However, it only includes three-phonon scattering due to the assumption that four-phonon and higher-order processes are generally unimportant. Also, directly evaluating the scattering rates of four-phonon and higher-order processes has been a long-standing challenge. In this work, however, we have developed a formalism to explicitly determine quantum mechanical scattering probability matrices for four-phonon scattering in the full Brillouin Zone, and by mitigating the computational challenge we have directly calculated four-phonon scattering rates. We find that four-phonon scattering rates are comparable to three-phonon scattering rates at medium and high temperatures, and they increase quadratically with temperature. As a consequence, $\kappa$ of Lennard-Jones argon is reduced by more than 60\% at 80 K when four-phonon scattering is included. Also, in less anharmonic materials -- diamond, silicon, and germanium, $\kappa$ is still reduced considerably at high temperature by four-phonon scattering by using the classical Tersoff potentials. Also, the thermal conductivity of optical phonons is dominated by the fourth and higher orders phonon scattering even at low temperature.
\end{abstract}

\maketitle

\section{Introduction}

Thermal transport in semiconductors and dielectrics is determined by phonon scattering processes. Intrinsic phonon-phonon scattering includes three-phonon, four-phonon and higher-order phonon processes. The scattering rate $\tau^{-1}$, reciprocal of phonon relaxation time $\tau$, is essential in predicting lattice thermal conductivity $\kappa$ based on the Boltzmann transport equation (BTE)\,\cite{Srivastava_book, Broido2007apl} \begin{equation} \kappa_z=\frac{1}{V}\sum_\lambda v_{z,\lambda}^2 c_\lambda\tau_\lambda, \label{eq_k_BTE}
\end{equation}where $V$ is the volume, $\lambda\equiv(\mathbf{k},j)$ specifies a phonon mode with wave vector $\mathbf{k}$ and dispersion branch $j$, $v_z$ is phonon group velocity projection along the transport $z$ direction, and $c_\lambda$ is phonon specific heat per mode. Starting from the third-order anharmonic Hamiltonian and the Fermi's golden rule (FGR), Maradudin \textit{et al.}\,\cite{Maradudin1962prb, Maradudin1962pss} proposed an anharmonic lattice dynamics (ALD) method to predict intrinsic three-phonon scattering rates in solids. Debernardi \textit{et al.}\,\cite{Deb1995prl} performed the ALD calculations based on density functional theory (DFT) to obtain $\tau_\lambda$. Recently significant advances have been achieved by Broido \textit{et al.} by combining ALD and BTE to predict $\kappa$\,\cite{Broido2007apl}. The ALD method based on first-principles force constants or classical interatomic potentials have since been extensively used\,\cite{Turney2009prb1, Lindsay2009prb, Esf2011prb, Lindsay2012prl}. A recent review on this topic can be found in Ref.\,\cite{Feng2014Jn}. However, the current ALD method is limited to evaluating three-phonon scattering rates and does not capture four and higher-order scatterings due to the challenge, and hence its accuracy is limited to relatively low temperatures, typically far from the melting point. For example, at 1000 K the lattice thermal conductivity of Si predicted by only considering three-phonon scattering is $\sim$41 W/mK\,\cite{Esf2011prb}, which is higher than the experimental value $\sim$30 W/m-K\,\cite{Glass1964PR}.

Although the study of four-phonon scattering has a long history, it was limited to the qualitative interpretation of the experimental data\,\cite{Joshi1970prb, Ecsedy1977prb}. Recently, Lindsay \textit{et al.} examined the phase space for four-phonon scattering processes \cite{Lindsay2008jp}. Turney \textit{et al.} have discussed the higher-order anharmonicity of the interatomic potential in argon, by comparing the three-phonon scattering rates obtained by the ALD method to the total phonon scattering rates obtained by molecular dynamics (MD) and normal mode analysis (NMA)\,\cite{Turney2009prb1}. Sapna and Singh\,\cite{Sapna2013mplb} estimated the four-phonon scattering rates in carbon nanotubes using an analytical model involving approximations such as the Callaway model, the Debye model, etc. Although NMA can predict the total scattering rates, it cannot separate three-phonon and higher-order phonon processes and does not provide scattering probability of each individual scattering process\,\cite{Feng2014Jn}. Therefore, a direct and rigorous calculation of four-phonon scattering rates in the ALD framework is of great significance for a better understanding of phonon transport and a more accurate prediction of $\kappa$.

In this work, we derive an ALD formalism for four-phonon scattering by extending the derivation of Maradudin \textit{et al.}\,\cite{Maradudin1962prb}. Bulk argon, a strongly anharmonic material, is used as a benchmark material to demonstrate the approach and the importance of four-phonon scattering in thermal transport. This is followed by the study of three less anharmonic materials -- bulk diamond, silicon and germanium. The accuracy of our calculations are demonstrated by the agreement of the scattering rates and lattice thermal conductivities between ALD (with four-phonon scattering included) and MD. Comparison is also made to experiment when appropriate. An agreement between our prediction and experimental results is also presented.

\section{Derivation of four-phonon scattering rate}

\label{Sec_formula}

The Hamiltonian of crystals can be written as the summation of the harmonic and anharmonic parts based on perturbation theory\,\cite{Maradudin1962prb,Ziman_book} 
\begin{equation}
	\hat{H} = \hat{H}_0 + \hat{H}_3 + \hat{H}_4 + \cdots,
\end{equation}
where the harmonic part $\hat{H}_0$, first-order perturbation $\hat{H}_3$ and second-order perturbation $\hat{H}_4$ are\,\cite{Maradudin1962prb}
\begin{eqnarray}
	&&\hat{H}_0=\sum_\lambda \hbar\omega_\lambda (a_\lambda^\dagger a_\lambda +1/2), \\
	&&\hat{H}_3 = \sum_{\lambda \lambda_1 \lambda_2} H_{\lambda\lambda_1\lambda_2}^{(3)}\left(a_{-\lambda}^\dagger + a_\lambda\right) (a_{-\lambda_1}^\dagger + a_{\lambda_1}) (a_{-\lambda_2}^\dagger + a_{\lambda_2}), \\
	&&\hat{H}_4 = \sum_{\lambda \lambda_1 \lambda_2 \lambda_3} H_{\lambda\lambda_1\lambda_2\lambda_3}^{(4)}(a_{-\lambda}^\dagger + a_\lambda) (a_{-\lambda_1}^\dagger + a_{\lambda_1}) (a_{-\lambda_2}^\dagger + a_{\lambda_2}) (a_{-\lambda_3}^\dagger + a_{\lambda_3}),
\end{eqnarray}
respectively. Here $a_\lambda^\dagger$ and $a_\lambda$ are the creation and annihilation operators with $a_\lambda^\dagger|n_\lambda\rangle=\sqrt{n_\lambda+1}|n_\lambda+1\rangle$ and $a_\lambda|n_\lambda\rangle=\sqrt{n_\lambda}|n_\lambda-1\rangle$ respectively. $\omega_\lambda$ is the angular frequency of the phonon mode $\lambda$. The expressions for $H_{\lambda\lambda_1\lambda_2}^{(3)}$ and $H_{\lambda\lambda_1\lambda_2\lambda_3}^{(4)}$ given in Ref.\,\cite{Maradudin1962prb} are
\begin{equation}
	H_{\lambda\lambda_1\lambda_2}^{(3)} = \frac{\hbar^{3/2}}{2^{3/2}\times 6N^{1/2}} \Delta_{\mathbf{k}\!+\!\mathbf{k}_1\!+\!\mathbf{k}_2,\mathbf{R}} \frac{V_{\lambda\lambda_1\lambda_2}^{(3)}} {\sqrt{\omega_\lambda\omega_{\lambda_1}\omega_{\lambda_2}}},
\end{equation}
\begin{equation}
	H_{\lambda\lambda_1\lambda_2\lambda_3}^{(4)} = \frac{\hbar^{2}}{2^{2}\times 24N} \Delta_{\mathbf{k}\!+\!\mathbf{k}_1\!+\!\mathbf{k}_2\!+\!\mathbf{k}_3,\mathbf{R}} \frac{V_{\lambda\lambda_1\lambda_2\lambda_3}^{(4)}} {\sqrt{\omega_\lambda\omega_{\lambda_1}\omega_{\lambda_2}\omega_{\lambda_3}}},
\end{equation}
\begin{equation}
	\label{eq_V3_H}
	V_{\lambda\lambda_1\lambda_2}^{(3)}\!=\!\sum_{b\!,l_1b_1\!,l_2b_2}\sum_{\alpha\alpha_1\!\alpha_2}\Phi_{0b\!,l_1b_1\!,l_2b_2}^{\alpha\alpha_1\alpha_2}\frac{e_{\alpha b}^\lambda e_{\alpha_1 b_1}^{\lambda_1} e_{\alpha_2 b_2}^{\lambda_2}}{\sqrt{\bar{m}_b \bar{m}_{b_1} \bar{m}_{b_2}}} e^{ i \mathbf{k_1}\cdot\mathbf{r}_{l_1}} e^{ i \mathbf{k_2}\cdot \mathbf{r}_{l_2}},
\end{equation}
\begin{equation}
	\label{eq_V4_H}
	V_{\lambda\lambda_1\lambda_2\lambda_3}^{(4)}=\sum_{b,l_1b_1,l_2b_2,l_3b_3}\sum_{\alpha\alpha_1\alpha_2\alpha_3}\Phi_{0b,l_1b_1,l_2b_2,l_3b_3}^{\alpha\alpha_1\alpha_2\alpha_3}\frac{e_{\alpha b}^\lambda e_{\alpha_1 b_1}^{\lambda_1} e_{\alpha_2 b_2}^{\lambda_2} e_{\alpha_3 b_3}^{\lambda_3}}{\sqrt{\bar{m}_b \bar{m}_{b_1} \bar{m}_{b_2} \bar{m}_{b_3}}} e^{ i \mathbf{k_1}\cdot \mathbf{r}_{l_1}} e^{ i \mathbf{k_2}\cdot \mathbf{r}_{l_2}} e^{ i \mathbf{k_3}\cdot \mathbf{r}_{l_3}},
\end{equation}
where $N$ is the total number of $\mathbf{k}$ points. $\mathbf{R}$ is a reciprocal lattice vector. The Kronecker deltas $\Delta_{\mathbf{k}+\mathbf{k}_1+\mathbf{k}_2,\mathbf{R}}$ and $\Delta_{\mathbf{k}+\mathbf{k}_1+\mathbf{k}_2\!+\mathbf{k}_3,\mathbf{R}}$ describe the momentum selection rule and have the property that $\Delta_{m,n}= 1$ (if $m=n$), or 0 (if $m\neq n$). $l$, $b$, and $\alpha$ label the indexes of the unit cells, basis atoms, and ($x$,$y$,$z$) directions, respectively. $\Phi_{0b\!,l_1b_1\!,l_2b_2}^{\alpha\alpha_1\alpha_2}$ and $\Phi_{0b,l_1b_1,l_2b_2,l_3b_3}^{\alpha\alpha_1\alpha_2\alpha_3}$ are the third- and fourth-order force constants, respectively.  $e$ is the phonon eigenvector. $\bar{m}_b$ is the average atomic mass at the lattice site $b$.

\begin{figure}[tbph]
	\centering
	\includegraphics[width= 3.3in]{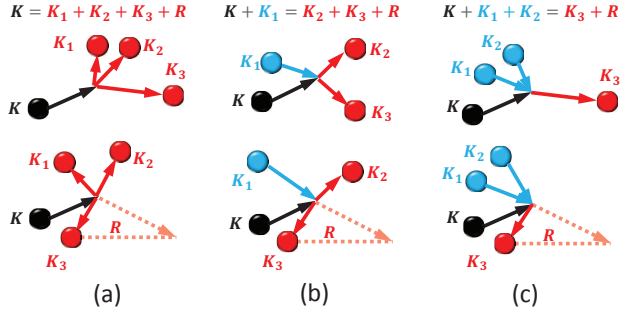}
	\caption{ The sketches of four-phonon scattering processes. (a)-(c) The splitting, redistribution, and combination processes, respectively. Each category contains $N$ processes ($\mathbf{R}=\mathbf{0}$) and $U$ processes ($\mathbf{R}\neq\mathbf{0}$).}\label{fig_scattering}
\end{figure}

Considering a three-phonon process $\lambda\rightarrow\lambda_1+ \lambda_2$, for example, the initial state is $|i\rangle=|n_\lambda+1,n_{\lambda_1},n_{\lambda_2}\rangle$ and the final state is $|f\rangle=|n_\lambda,n_{\lambda_1}+1,n_{\lambda_2}+1\rangle$. Based on FGR, the transition probability from $|i\rangle$ to $|f\rangle$ is proportional to 
\begin{equation}
	\frac{2\pi}{\hbar}\left|\langle f|\hat{H}_3|i\rangle\right|^2\delta(E_i-E_f) \sim \left|\sqrt{n_\lambda}\sqrt{1\!+\!n_{\lambda_1}}\sqrt{\!1\!+\!n_{\lambda_2}\!}\right|^2\cdot\left|H_{\lambda\lambda_1\lambda_2}^{(3)}\right|^2 \sim n_\lambda (1\!+\!n_{\lambda_1})(\!1\!+\!n_{\lambda_2}\!)\left|H_{\lambda\lambda_1\lambda_2}^{(3)}\right|^2.
\end{equation}
Similarly the transition probability of the process $\lambda\leftarrow\lambda_1+ \lambda_2$ is proportional to 
\begin{equation}
	\frac{2\pi}{\hbar}\left|\langle i|\hat{H}_3|f\rangle\right|^2\delta(E_i-E_f) \sim \left|\sqrt{1+n_\lambda}\sqrt{n_{\lambda_1}}\sqrt{\!n_{\lambda_2}\!}\right|^2\cdot\left|H_{\lambda\lambda_1\lambda_2}^{(3)}\right|^2 \sim (\!1\!+\!n_\lambda\!)n_{\lambda_1}n_{\lambda_2}\left|H_{\lambda\lambda_1\lambda_2}^{(3)}\right|^2.
\end{equation}
The time rate of the occupation number change of the mode $\lambda$ due to  three-phonon\,\cite{Maradudin1962prb, Feng2014Jn, Klemens_book, Ziman_book, Kaviany_book} and four-phonon scattering (Fig.\,\ref{fig_scattering}) can be written as
\begin{eqnarray}
	\label{BTE2}
	\frac{\partial n_\lambda}{\partial t} |_s = 
	&& -\sum_{\lambda_1\lambda_2}\left\{\frac{1}{2}\big[n_\lambda (1\!+\!n_{\lambda_1})(\!1\!+\!n_{\lambda_2}\!)\!-\!(\!1\!+\!n_\lambda\!)n_{\lambda_1}n_{\lambda_2}\big]\mathcal{L}_- \!+\! \big[n_\lambda n_{\lambda_1}(\!1\!+\!n_{\lambda_2}\!)\!-\!(\!1\!+\!n_\lambda\!)(\!1\!+\!n_{\lambda_1}\!)n_{\lambda_2}\big]\mathcal{L}_+\right\} \nonumber \\ 
	& & -\sum_{\lambda_1\lambda_2\lambda_3}\bigg\{ \frac{1}{6}\big[n_\lambda (\!1\!+\!n_{\lambda_1}\!)(\!1\!+\!n_{\lambda_2}\!)(\!1\!+\!n_{\lambda_3}\!)\!-\!(\!1\!+\!n_\lambda\!)n_{\lambda_1}n_{\lambda_2}n_{\lambda_3}\big]\mathcal{L}_{--} \nonumber \\ 
	& & \ \ \ \ \ \ \ \ \ \ \ \ \!+ \frac{1}{2}\big[n_\lambda n_{\lambda_1}(1\!+\!n_{\lambda_2})(\!1\!+\!n_{\lambda_3}\!)\!-\!(\!1\!+\!n_\lambda\!)(\!1\!+\!n_{\lambda_1}\!)n_{\lambda_2}n_{\lambda_3}\big]\mathcal{L}_{+-} \nonumber \\ 
	& & \ \ \ \ \ \ \ \ \ \ \ \ \!+ \frac{1}{2}\big[n_\lambda n_{\lambda_1}n_{\lambda_2}(\!1\!+\!n_{\lambda_3}\!)\!-\!(\!1\!+\!n_\lambda\!)(\!1\!+\!n_{\lambda_1}\!)(1\!+\!n_{\lambda_2})n_{\lambda_3}\big]\mathcal{L}_{++} \bigg\} 
\end{eqnarray}
The first summation on the right hand side represents the three-phonon scattering rate of the mode $\lambda$, with the first term accounting for the splitting process $\lambda\rightarrow\lambda_1+ \lambda_2$ and the second the combination process $\lambda+\lambda_1\rightarrow \lambda_2$. The physical meaning of the first term is the difference between the transition rates of $\lambda\rightarrow\lambda_1+ \lambda_2$ and $\lambda\leftarrow\lambda_1+ \lambda_2$, and thus indicates the decay rate of $n_\lambda$ due to the splitting process.  Similarly, the second term illustrates the transition rate difference between $\lambda+\lambda_1\rightarrow \lambda_2$ and $\lambda+\lambda_1\leftarrow \lambda_2$, indicating the decay rate of $n_\lambda$ due to the combination process. $\mathcal{L}_\pm$ contains the information of the intrinsic transition probability and the transition selection rules for energy and momentum, $\omega_\lambda\pm\omega_{\lambda_1}-\omega_{\lambda_2}=0$ and $\mathbf{k}\pm\mathbf{k}_1-\mathbf{k}_2=\mathbf{R}$, with $\mathbf{R}=0$ implying the normal ($N$) process and $\mathbf{R}\neq 0$ the Umklapp ($U$) process. The second summation accounts for the four-phonon scattering of the mode $\lambda$, with the first parentheses representing the process $\lambda\rightarrow \lambda_1 + \lambda_2+\lambda_3$, the second the process $\lambda+ \lambda_1 \rightarrow \lambda_2+\lambda_3$, and the third $\lambda+ \lambda_1 + \lambda_2\rightarrow\lambda_3$. Similarly, $\mathcal{L}_{\pm\pm}$ accounts for the transition probabilities and the selection rules $\omega_\lambda\pm\omega_{\lambda_1}\pm\omega_{\lambda_2}-\omega_{\lambda_3}=0$ and $\mathbf{k}\pm\mathbf{k}_1\pm\mathbf{k}_2-\mathbf{k}_3=\mathbf{R}$. The minus sign before each scattering term indicates that the perturbation $n_\lambda'$ to the equilibrium Bose-Einstein distribution  $n_\lambda^0$ is decreasing with time, i.e., the phonon distribution tends to recover its equilibrium state, due to the scattering. The factors 1/6 and 1/2 in Eq.\,(\ref{BTE2}) account for the sixfold count and double count in the summation, respectively.  

In the single mode relaxation time approximation (SMRTA)\,\cite{Kaviany_book, Feng2014Jn}, the mode $\lambda$ is suddenly stimulated to an excited state and has the occupation number
\begin{equation}
	n_\lambda = n_\lambda^0+n_\lambda',\label{eq_n}
\end{equation}
while other modes stay in equilibrium, i.e.,
\begin{eqnarray}
	\label{n_Iterative}
	n_{\lambda_1} &=& n_{\lambda_1}^0, \label{eq_nprime} \\
	n_{\lambda_2} &=& n_{\lambda_2}^0, \label{eq_n2prime} \\
	n_{\lambda_3} &=& n_{\lambda_3}^0. \label{eq_n3prime}
\end{eqnarray}
By substituting Eqs.\,(\ref{eq_n})-(\ref{eq_n3prime}) into Eq.\,(\ref{BTE2}) and using the fact that
\begin{eqnarray}
	\lambda\!\rightarrow\!\lambda_1\!+\!\lambda_2\!&:&
	n_\lambda^0(1+n_{\lambda_1}^0)(1+n_{\lambda_2}^0)-(1+n_\lambda^0)n_{\lambda_1}^0 n_{\lambda_2}^0 =0 \label{eq_math_0}\\
	\lambda\!+\!\lambda_1\!\rightarrow\!\lambda_2\!&:& 
	n_\lambda^0n_{\lambda_1}^0(1+n_{\lambda_2}^0) - (1+n_\lambda^0)(1+n_{\lambda_1}^0)n_{\lambda_2}^0 =0 \label{eq_math_2}\\
	\lambda\!\rightarrow\!\lambda_1\!+\!\lambda_2\!+\!\lambda_3\!&:&
	n_\lambda^0(1+n_{\lambda_1}^0)(1+n_{\lambda_2}^0)(1+n_{\lambda_3}^0)-(1+n_{\lambda}^0)n_{\lambda_1}^0 n_{\lambda_2}^0 
	n_{\lambda_3}^0 =0\\
	\lambda\!+\!\lambda_1\!\rightarrow\!\lambda_2\!+\!\lambda_3\!&:&
	n_\lambda^0n_{\lambda_1}^0(1+n_{\lambda_2}^0)(1+n_{\lambda_3}^0)-(1+n_{\lambda}^0)(1+n_{\lambda_1}^0) n_{\lambda_2}^0 n_{\lambda_3}^0 =0\\
	\lambda\!+\!\lambda_1\!+\!\lambda_2\!\rightarrow\!\lambda_3\!&:&
	n_\lambda^0n_{\lambda_1}^0 n_{\lambda_2}^0(1+n_{\lambda_3}^0)-(1+n_{\lambda}^0)(1+n_{\lambda_1}^0)(1+n_{\lambda_2}^0) n_{\lambda_3}^0 =0
\end{eqnarray}
and the fact
\begin{eqnarray}
	\lambda\!\rightarrow\!\lambda_1\!+\!\lambda_2\!&:& (1+n_{\lambda_1}^0)(1+n_{\lambda_2}^0)-n_{\lambda_1}^0 n_{\lambda_2}^0 = \frac{n_{\lambda_1}^0 n_{\lambda_2}^0}{n_{\lambda}^0} = 1+n_{\lambda_1}^0+n_{\lambda_2}^0, \label{eq_math1}\\
	\lambda\!+\!\lambda_1\!\rightarrow\!\lambda_2\!&:& n_{\lambda_1}^0(1+n_{\lambda_2}^0) - (1+n_{\lambda_1}^0)n_{\lambda_2}^0 = \frac{(1+n_{\lambda_1}^0)n_{\lambda_2}^0}{n_{\lambda}^0} = n_{\lambda_1}^0- n_{\lambda_2}^0, \label{eq_math2}\\
	\lambda\!\rightarrow\!\lambda_1\!+\!\lambda_2\!+\!\lambda_3\!&:&
	(1+n_{\lambda_1}^0)(1+n_{\lambda_2}^0)(1+n_{\lambda_3}^0)-n_{\lambda_1}^0 n_{\lambda_2}^0 
	n_{\lambda_3}^0 = \frac{n_{\lambda_1}^0 n_{\lambda_2}^0 n_{\lambda_3}^0}{n_{\lambda}^0},\label{eq_math3}\\
	\lambda\!+\!\lambda_1\!\rightarrow\!\lambda_2\!+\!\lambda_3\!&:&
	n_{\lambda_1}^0(1+n_{\lambda_2}^0)(1+n_{\lambda_3}^0)-(1+n_{\lambda_1}^0) n_{\lambda_2}^0 n_{\lambda_3}^0 = \frac{(1+n_{\lambda_1}^0) n_{\lambda_2}^0 n_{\lambda_3}^0}{n_{\lambda}^0}, \label{eq_math4}\\
	\lambda\!+\!\lambda_1\!+\!\lambda_2\!\rightarrow\!\lambda_3\!&:&
	n_{\lambda_1}^0 n_{\lambda_2}^0(1+n_{\lambda_3}^0)-(1+n_{\lambda_1}^0)(1+n_{\lambda_2}^0) n_{\lambda_3}^0 = \frac{(1+n_{\lambda_1}^0) (1+n_{\lambda_2}^0) n_{\lambda_3}^0}{n_{\lambda}^0},\label{eq_math5}
\end{eqnarray}
Eq.\,(\ref{BTE2}) is reduced to
\begin{eqnarray}
	\label{BTE3}
	\frac{\partial n_\lambda'}{\partial t} |_s =& & -n_\lambda'\sum_{\lambda_1\lambda_2}\left\{   \frac{1}{2}(1+n_{\lambda_1}^0+n_{\lambda_2}^0)\mathcal{L}_- \!+\! (n_{\lambda_1}^0- n_{\lambda_2}^0)\mathcal{L}_+ \right\} \nonumber\\
	& & -n_\lambda'\sum_{\lambda_1\lambda_2\lambda_3}\bigg\{ \frac{1}{6} \frac{n_{\lambda_1}^0 n_{\lambda_2}^0 n_{\lambda_3}^0}{n_{\lambda}^0} \mathcal{L}_{--} \nonumber \!+ \frac{1}{2} \frac{(1+n_{\lambda_1}^0) n_{\lambda_2}^0 n_{\lambda_3}^0}{n_{\lambda}^0} \mathcal{L}_{+-} \nonumber \!+ \frac{1}{2}\frac{(1+n_{\lambda_1}^0) (1+n_{\lambda_2}^0) n_{\lambda_3}^0}{n_{\lambda}^0} \mathcal{L}_{++} \bigg\} \nonumber \\ 
	=& & -n_\lambda'(\tau_{3,\lambda}^{-1}+ \tau_{4,\lambda}^{-1}),
\end{eqnarray}
where $\tau_{3,\lambda}^{-1}$ and $\tau_{4,\lambda}^{-1}$ are
\begin{equation}
\label{tau3}
  \tau_{3,\lambda}^{-1}= \sum_{\lambda_1\lambda_2}\left\{   \frac{1}{2}(1+n_{\lambda_1}^0+n_{\lambda_2}^0)\mathcal{L}_- \!+\! (n_{\lambda_1}^0- n_{\lambda_2}^0)\mathcal{L}_+ \right\} ,
\end{equation}

\begin{equation}
\label{tau4}
\tau_{4,\lambda}^{-1}=  \sum_{\lambda_1\lambda_2\lambda_3}\bigg\{ \frac{1}{6} \frac{n_{\lambda_1}^0 n_{\lambda_2}^0 n_{\lambda_3}^0}{n_{\lambda}^0} \mathcal{L}_{--} \!+ \frac{1}{2} \frac{(1+n_{\lambda_1}^0) n_{\lambda_2}^0 n_{\lambda_3}^0}{n_{\lambda}^0} \mathcal{L}_{+-} \!+ \frac{1}{2}\frac{(1+n_{\lambda_1}^0) (1+n_{\lambda_2}^0) n_{\lambda_3}^0}{n_{\lambda}^0} \mathcal{L}_{++} \bigg\}.
\end{equation}
Thus, the scattering rate based on the SMRTA is 
\begin{equation}
	\label{eq_solution_SMRTA}
	\tau_\lambda^{-1}= \tau_{3,\lambda}^{-1}+ \tau_{4,\lambda}^{-1}.
\end{equation}
The exact solution to BTE beyond the SMRTA including four-phonon scattering is quite complicated and thus will be presented in our subsequent work. Since the focus of this paper is on the importance of four-phonon scattering compared to the three-phonon scattering, the SMRTA is enough to demonstrate the features. 

Equations (\ref{eq_math_0})-(\ref{eq_math5}) are derived based on the energy conservation law. For example, Eqs.\,(\ref{eq_math_0}) and (\ref{eq_math1}) are derived by substituting the $\omega$ of the Bose-Einstein distribution $e^{\hbar\omega/k_BT}=1+1/n_\lambda^0$ into the energy conservation (selection rule) $\omega_\lambda=\omega_{\lambda_1}+\omega_{\lambda_2}$, giving the relation $1+1/n_\lambda^0=(1+1/n_{\lambda_1}^0)(1+1/n_{\lambda_2}^0)$.

The expressions for $\mathcal{L}_\pm$ and $\mathcal{L}_{\pm\pm}$ are given by FGR,
\begin{eqnarray}
	\label{Lpm}
	\mathcal{L}_\pm&=& 18*2\frac{2\pi}{\hbar}\left|H_{\lambda\lambda_1\lambda_2}^{(3)}\right|^2\delta(E_i-E_f)\label{eq_Lpm1}\\ &=&\frac{\pi\hbar}{4N}\!\left|V_{\pm}^{\!(3)\!}\right|^2\! \Delta_\pm \frac{\delta(\omega_\lambda\!\pm\!\omega_{\lambda_1}\!-\!\omega_{\lambda_2})}{\omega_\lambda\omega_{\lambda_1}\omega_{\lambda_2}}\label{Lpm2},
\end{eqnarray}
\begin{eqnarray}
	\label{Lpmpm}
	\mathcal{L}_{\pm\pm}&=& 96*2\frac{2\pi}{\hbar}\left|H_{\lambda\lambda_1\lambda_2\lambda_3}^{(4)}\right|^2\delta(E_i-E_f)\label{eq_Lpmpm1}\\
	&=& \frac{\pi\hbar}{4N} \frac{\hbar}{2N} \!\left|V_{\pm\pm}^{\!(4)\!}\right|^2\! \Delta_{\pm\pm} \frac{\delta(\omega_\lambda\pm\omega_{\lambda_1}\pm\omega_{\lambda_2}-\omega_{\lambda_3})}{\omega_\lambda\omega_{\lambda_1}\omega_{\lambda_2}\omega_{\lambda_3}}\label{Lpmpm2},
\end{eqnarray}
where $V_{\pm}^{(3)}$ and $V_{\pm\pm}^{(4)}$ are
\begin{equation}
	\label{eq_V3}
	V_{\pm}^{(3)}\!=\!\sum_{b\!,l_1b_1\!,l_2b_2}\sum_{\alpha\alpha_1\!\alpha_2}\Phi_{0b\!,l_1b_1\!,l_2b_2}^{\alpha\alpha_1\alpha_2}\frac{e_{\alpha b}^\lambda e_{\alpha_1 b_1}^{\pm\lambda_1} e_{\alpha_2 b_2}^{-\lambda_2}}{\sqrt{\bar{m}_b \bar{m}_{b_1} \bar{m}_{b_2}}} e^{\pm i \mathbf{k_1}\!\cdot \!\mathbf{r}_{l_1}} e^{- i \mathbf{k_2}\!\cdot\! \mathbf{r}_{l_2}},
\end{equation}
\begin{equation}
	\label{eq_V4}
	V_{\pm\pm}^{(4)}=\sum_{b,l_1b_1,l_2b_2,l_3b_3}\sum_{\alpha\alpha_1\alpha_2\alpha_3}\Phi_{0b,l_1b_1,l_2b_2,l_3b_3}^{\alpha\alpha_1\alpha_2\alpha_3}\frac{e_{\alpha b}^\lambda e_{\alpha_1 b_1}^{\pm\lambda_1} e_{\alpha_2 b_2}^{\pm\lambda_2} e_{\alpha_3 b_3}^{-\lambda_3}}{\sqrt{\bar{m}_b \bar{m}_{b_1} \bar{m}_{b_2} \bar{m}_{b_3}}} e^{\pm i \mathbf{k_1}\cdot \mathbf{r}_{l_1}} e^{\pm i \mathbf{k_2}\cdot \mathbf{r}_{l_2}} e^{- i \mathbf{k_3}\cdot \mathbf{r}_{l_3}}.
\end{equation}
In Eq.\,(\ref{eq_Lpm1}), the factor 18 accounts for the topologically equivalent pairing schemes that were explained in Ref.\,\cite{Maradudin1962prb}. Analogously, the factor 96 in Eq. (\ref{eq_Lpmpm1}) comes from the fact that in Fig.\,5 of Ref.\,\cite{Maradudin1962prb} the phonon $\lambda$ can pair with any of the four phonons at the lower vertex, and the $\mathbf{k}j'$ can pair with any four at the upper vertex, while the three remaining phonons at lower vertex can pair with the three remaining phonons at the upper vertex in six ways. In both Eqs. (\ref{eq_Lpm1}) and (\ref{eq_Lpmpm1}), the factor 2 in front of $\frac{2\pi}{\hbar}$ accounts for the difference between scattering rate and self-energy linewidth. The delta function $\delta(E)$ is replaced by $\delta(\omega)/\hbar$. The Kronecker deltas $\Delta_\pm$ and $\Delta_{\pm\pm}$ are short for $\Delta_{\mathbf{k}\pm\mathbf{k}_1-\mathbf{k}_2,\mathbf{R}}$ and $\Delta_{\mathbf{k}\pm\mathbf{k}_1\pm\mathbf{k}_2\!-\mathbf{k}_3,\mathbf{R}}$, respectively.

\section{mitigate the computational cost}


We use the central difference method to obtain the second-, third-, and fourth-order IFCs as listed below: 
\begin{equation}
\Phi_{at_1,at_2}^{\alpha_1\alpha_2}=\frac{1}{(2\Delta)^2}\sum_{s_1,s_2}^{-1,1}s_1s_2E(r_{at_1}^{\alpha_1}+s_1\Delta,r_{at_2}^{\alpha_2}+s_2\Delta),
\end{equation}
\begin{equation}
\Phi_{at_1,at_2,at_3}^{\alpha_1\alpha_2\alpha_3}=\frac{1}{(2\Delta)^3}\sum_{s_1,s_2,s_3}^{-1,1}s_1s_2s_3E(r_{at_1}^{\alpha_1}+s_1\Delta,r_{at_2}^{\alpha_2}+s_2\Delta,r_{at_3}^{\alpha_3}+s_3\Delta),
\end{equation}
\begin{equation}
\Phi_{at_1,at_2,at_3,at_4}^{\alpha_1\alpha_2\alpha_3\alpha_4}=\frac{1}{(2\Delta)^4}\sum_{s_1,s_2,s_3,s_4}^{-1,1}s_1s_2s_3s_4E(r_{at_1}^{\alpha_1}+s_1\Delta,r_{at_2}^{\alpha_2}+s_2\Delta,r_{at_3}^{\alpha_3}+s_3\Delta,r_{at_4}^{\alpha_4}+s_4\Delta).
\end{equation}
Here $r_{at}^{\alpha}$ represents the $\alpha$ component of the equilibrium position of the atom $at$. $\Delta$ is a all displacement. $E$ is the energy. In the central difference method, the energy derivatives at the position $\mathbf{r}$ are obtained by the finite differences between $E(r^{\alpha}+\Delta)$ and $E(r^{\alpha}-\Delta)$, ($\alpha=x,y,z$), and thus $s_1$ \textit{et al.} take the two values ``+1'' and ``-1'', representing the plus and minus signs, respectively. The calculation of the $m$th-order IFC requires the energies of $(6Nn_b)^m$ atomic configurations.

The computational cost and the memory requirement of fourth-order ALD calculations by Eqs.(\ref{Lpmpm2}) and (\ref{eq_V4}) are 9$NN_cn_b^2$ times of those of third-order by Eqs.(\ref{Lpm2}) and (\ref{eq_V3}), where $n_b$ and $N_c$ are the total number of basis atoms in a primitive cell and the total number of primitive cells in the domain, respectively.
To obtain the phonon scattering rate, the computational cost needs to be reduced. In first-principles-based three-phonon scattering calculations, the most time consuming part is to obtain the IFCs, and symmetries were typically employed to reduce the computational cost\,\cite{Esfar2008prb, Li2012prb}. In contrast, in our classical potential-based four-phonon scattering calculation, the biggest challenge is the scattering matrices calculation rather than the IFCs calculation since the phase space allows $10^3$-$10^4$ orders larger amount of 4-phonon processes than 3-phonon processes for the $\mathbf{k}$ meshes studied in this work. Therefore, the essential works are 1) to reduce the number of IFCs, which directly affects the computational cost of each scattering matrix element of $V_\pm^{(3)}$ and $V_{\pm\pm}^{(4)}$, and 2) to reduce the dimensions of $V_\pm^{(3)}$ and $V_{\pm\pm}^{(4)}$ by excluding in advance the mode combinations that do not satisfy the momentum and energy selection rules. In the calculation of IFCs, even if every pair of atoms is within the potential cutoff radius, the force-constant is not necessary to be counted. For example, typically the value of a 3-IFC or 4-IFC does not depend on the choice of the small atomic displacement $\Delta$, if the value of a IFC is found to be near 0 and strongly depend on the displacement $\Delta$ (e.g., the value may vary by orders near 0 when $\Delta$ changes by 2 times, and it does not converge no matter what value the $\Delta$ is), this IFC is considered to be negligible compared to numerical accuracy. By testing sufficient cases and finding out the conditions that have such infinitesimal IFCs, one can exclude the atomic combinations that satisfy such conditions in advance. To validate the accuracy of the calculation after employing these optimizations, we have compared the results of three-phonon scattering rates before and after employing these optimizations. No difference was found within the recorded precision for the three-phonon scattering rates, indicating that those optimizations do not sacrifice the calculation accuracy. These optimizations may not be significant for three-phonon scattering calculations since the total computation cost is relatively low, however, they are essential for making the four-phonon calculations practical. Even after these optimizations, the $V_{\pm\pm}$ matrices for 4-phonon process may still largely exceed the maximum memory of computers. By separating the calculation into several steps and writing/reading the data to/from files, this problem can be solved. Last but not least, to predict thermal conductivity, only the phonon scattering rates in the irreducible BZ are needed to be calculated.

For argon, the Lennard-Jones potential\,\cite{Ashcroft_book} with a cutoff radius of 8.5 \AA{} is used to describe the interatomic interaction. The scattering rates are calculated on the mesh of $16\!\times\!16\!\times\!8$ $\mathbf{k}$-points in the Brillouin zone. For diamond, Si, and Ge, the Tersoff potential\,\cite{Tersoff1989} and a $16\!\times\!16\!\times\!16$ $\mathbf{k}$-mesh are used. The details of normal mode analysis and Green-Kubo method based on molecular dynamics are described in Appendix\,\ref{append}.


\section{Results}
\subsection{Benchmark on Lennard-Jones argon: Large four-phonon scattering rates}
\begin{figure}[tphb]
	\begin{center}
		\includegraphics[width= 0.4\linewidth]{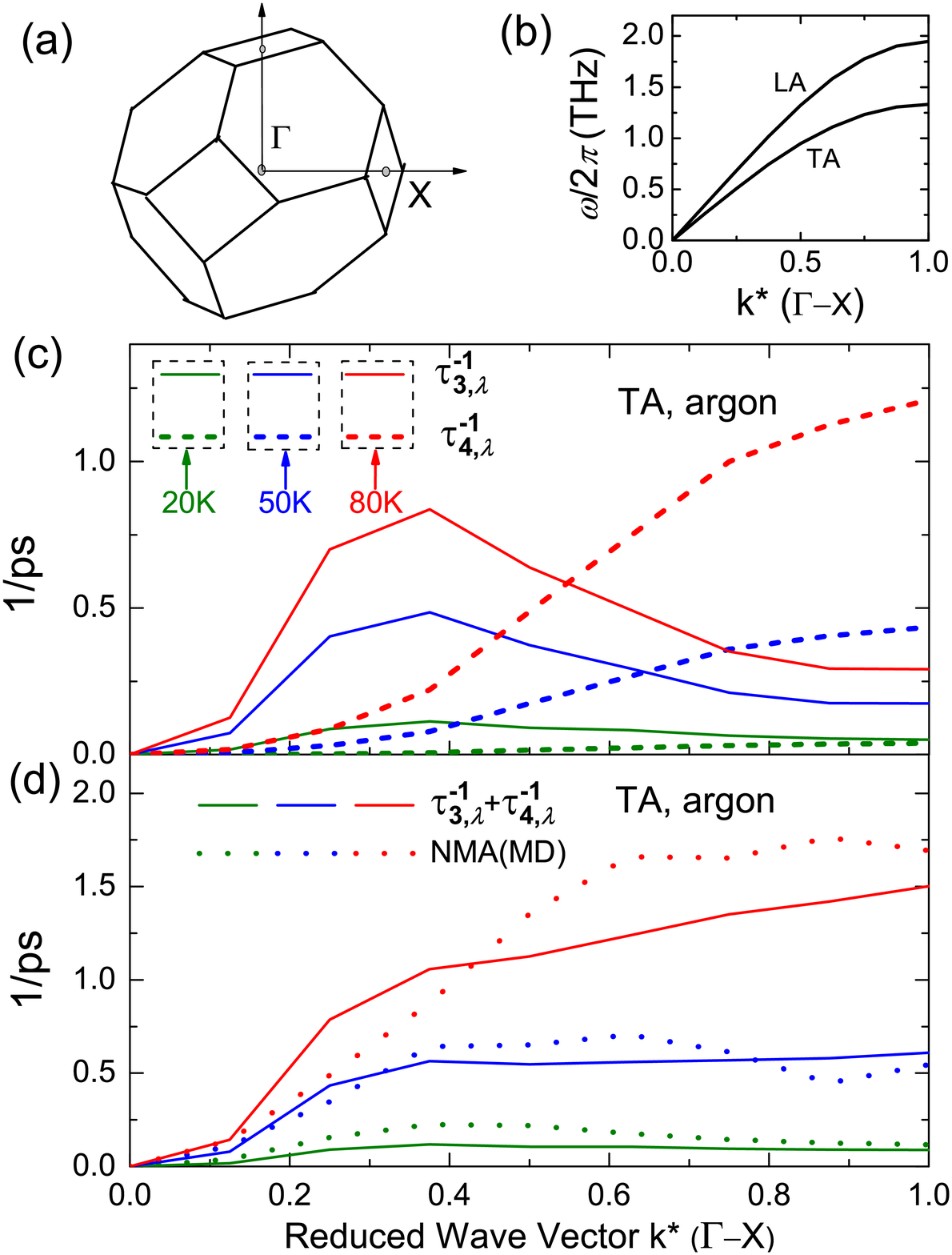}
	\end{center}
	\caption{(a) The Brillouin Zone for face-centered cubic structures (Ar, diamond, Si and Ge). (b) Dispersion relation of Ar from $\Gamma$ to X. (c) $\tau_{3,\lambda}^{-1}$ and $\tau_{4,\lambda}^{-1}$ of the TA branch with respect to the reduced wave vector ($\Gamma$-X) in argon at 20, 50, and 80 K, which are represented by different colors. (d) $\tau_{3,\lambda}^{-1}+\tau_{4,\lambda}^{-1}$ is compared to the linewidth $\tau_{{\rm NMA},\lambda}^{-1}$ predicted in frequency-domain NMA based on MD.}\label{TA_Ar}
\end{figure}

\begin{figure}[t]
	\centering
	\includegraphics[width= 3.4in]{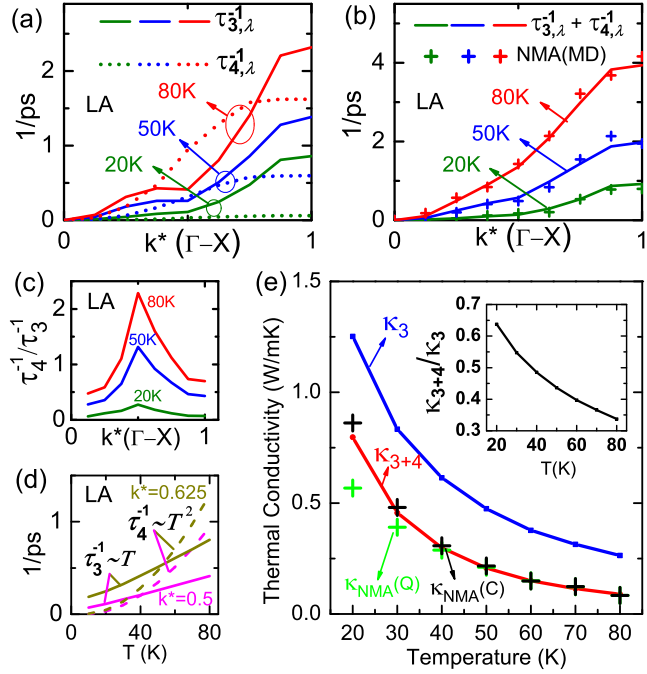}
	\caption{Phonon scattering rates and thermal conductivity of argon. (a) The $\tau_{3,\lambda}^{-1}$ and $\tau_{4,\lambda}^{-1}$ of the LA branch as a function of the reduced wave vector $\mathbf{k}^*$ from $\Gamma$ to X in argon at 20 K, 50 K and 80 K. (b) The summation of $\tau_{3,\lambda}^{-1}$ and $\tau_{4,\lambda}^{-1}$ is compared to the linewidth $\tau_{{\rm NMA},\lambda}^{-1}$ predicted in frequency-domain NMA based on MD simulations. (c) The relative importance of the four-phonon scattering rates, $\tau_{4,\lambda}^{-1}/\tau_{3,\lambda}^{-1}$, for the LA branch at 20 K, 50 K and 80 K. (d) The temperature dependences of $\tau_{3,\lambda}^{-1}\sim T$ and $\tau_{4,\lambda}^{-1}\sim T^2$ for the two modes $k\mathbf{k}^*$=(0.5,0,0) and $\mathbf{k}^*$=(0.625,0,0). (e) The $\kappa$ values of argon predicted from $\tau_{3,\lambda}^{-1}$, $\tau_{3,\lambda}^{-1}+\tau_{4,\lambda}^{-1}$, and $\tau_{{\rm NMA},\lambda}^{-1}$ as a function of temperature, with the inset showing the ratio of $\kappa_{3+4}/\kappa_{3}$. $\kappa_{\rm NMA}$(Q) and $\kappa_{\rm NMA}$(C) represent that the specific heat $c_\lambda$ in Eq.\,(\ref{eq_k_BTE}) is calculated by the quantum (Bose-Einstein) and classical (Boltzmann) phonon distributions, respectively. The phonon dispersion used in the calculation of $\kappa_{\rm NMA}$ is from lattice dynamics (LD) calculation, to be consistent with the $\kappa_3$ and $\kappa_{3+4}$ calculations.}\label{fig_argon_comp}
\end{figure}

Taking bulk argon as a benchmark material, which has been extensively studied\,\cite{Feng2014Jn, McGaughey2004prb, Turney2009prb1, Kab2007jap}, we have calculated the spectral scattering rates $\tau_{3,\lambda}^{-1}$ and $\tau_{4,\lambda}^{-1}$ as shown in Figs.\,\ref{TA_Ar} and\,\ref{fig_argon_comp}. Interestingly, we found that $\tau_{4,\lambda}^{-1}$ is comparable to $\tau_{3,\lambda}^{-1}$ at mid and high temperatures. To benchmark the accuracy of the calculation, we carried out MD simulations and frequency-domain NMA to probe the linewidth $\tau_{{\rm NMA},\lambda}^{-1}$ of the phonon spectral energy density, which includes the total scattering rates of all orders. It can be seen that $\tau_{3,\lambda}^{-1}+\tau_{4,\lambda}^{-1}$ agrees well with $\tau_{{\rm NMA},\lambda}^{-1}$ for both the TA and LA branches throughout the frequency and temperature range as shown in Figs.\,\ref{TA_Ar}(d) and \ref{fig_argon_comp}\,(b). In addition, the values of $\tau_{3,\lambda}^{-1}$ and $\tau_{{\rm NMA},\lambda}^{-1}$ agree well with those predicted by Turney \textit{et al.}\,\cite{Turney2009prb1} using ALD and the time-domain NMA, respectively.

The reason why $\tau_{4,\lambda}^{-1}$ is comparable to $\tau_{3,\lambda}^{-1}$ is that although each four-phonon process is in a higher-order and thus has a much lower scattering probability, the momentum and energy selection rules allow much greater number of four-phonon processes. For instance, each phonon mode participates in $\sim 10^3\!-\!10^4$ three-phonon processes while $\sim 10^7$ four-phonon processes for argon in a $16\!\times\!16\!\times\!8$ $\mathbf{k}$-mesh. In Fig.\,\ref{fig_argon_comp}\,(c), we show the relative importance of four-phonon scattering $\tau_{4,\lambda}^{-1}/\tau_{3,\lambda}^{-1}$ with respect to the reduced wave vector. The mid-frequency LA phonons have the highest $\tau_{4,\lambda}^{-1}$, since all the three types of four-phonon processes in Fig.\,\ref{fig_scattering} are allowed to happen.

Another important note is that at high temperatures $\tau_{3,\lambda}^{-1}$ increases linearly whereas $\tau_{4,\lambda}^{-1}$ quadratically with increasing temperature\,\cite{Joshi1970prb}, as shown in Fig.\,\ref{fig_argon_comp}\,(d). These temperature dependencies result from Eq.\,(\ref{tau3}) and Eq.\,(\ref{tau4}), which roughly indicate $\tau_{3,\lambda}^{-1}\sim n^0$ and $\tau_{4,\lambda}^{-1}\sim (n^0)^2$, leading to $\tau_{3,\lambda}^{-1}\sim T$ and $\tau_{4,\lambda}^{-1}\sim T^2$ since $n^0$ is proportional to $T$ at high temperatures.

The importance of four-phonon scattering in lattice thermal conductivity is studied by calculating $\kappa_3$, $\kappa_{3+4}$ and $\kappa_{\rm NMA}$ based on $\tau_{3,\lambda}^{-1}$, $\tau_{3,\lambda}^{-1}+\tau_{4,\lambda}^{-1}$ and $\tau_{{\rm NMA},\lambda}^{-1}$ respectively. For less computational cost, we use the isotropic assumption by taking the phonon modes in the $\Gamma-X$ direction to calculate the thermal conductivity. Equation (\ref{eq_k_BTE}) is converted to the continuous form\,\cite{Feng2014Jn} $\kappa_z=\frac{1}{(2\pi)^3}\sum_j\int c_\lambda v_{\lambda,z}^2 \tau_\lambda d\mathbf{k} =\frac{4\pi}{3}\frac{1}{(2\pi)^3}\sum_j \int c_\lambda v_\lambda^2 \tau_\lambda k^2 dk$ by taking the facts that $\sum_\mathbf{k}=\frac{V}{(2\pi)^3}\int d\mathbf{k}=\frac{V}{(2\pi)^3}\int 4\pi k^2 dk$ and that the integration of $|v_{\lambda,z}|^2$ gives $v_\lambda^2/3$. As shown in Fig.\,\ref{fig_argon_comp}\,(e), $\kappa_{3+4}$ and $\kappa_{\rm NMA}$ agree well with each other as well as those by Turney \textit{et al.}\,\cite{Turney2009prb1}. In contrast, $\kappa_3$ is considerably over-predicted especially at high temperatures. For a clearer insight, we plot the ratio of $\kappa_{3+4}/\kappa_3$ as a function of temperature in the inset. Four-phonon scattering reduces $\kappa$ of argon by 35\%-65\% at temperatures of 20-80 K. The results clearly demonstrate the importance of four-phonon scattering in thermal transport in a strongly anharmonic material or at high temperature. We note that $\kappa_{\rm NMA}$ is based on MD simulations which follow the classical (Boltzmann) distribution, while ALD calculations are based on quantum (Bose-Einstein) phonon distribution. Thus, the agreement between $\kappa_{3+4}$ and $\kappa_{\rm NMA}$ is better at higher temperatures where quantum physics is closer to classical physics. For low temperatures we did not attempt to replace the Bose-Einstein distribution in the ALD formula with the Boltzmann distribution in order to compare with the MD results, since this approach is not exact (See Sec.\,\ref{Sec_Boltz} for details).

\subsection{Significant four-phonon scattering rates in diamond, Si and Ge at high temperatures}

\begin{figure}[tbph]
	\includegraphics[width= 0.45\linewidth]{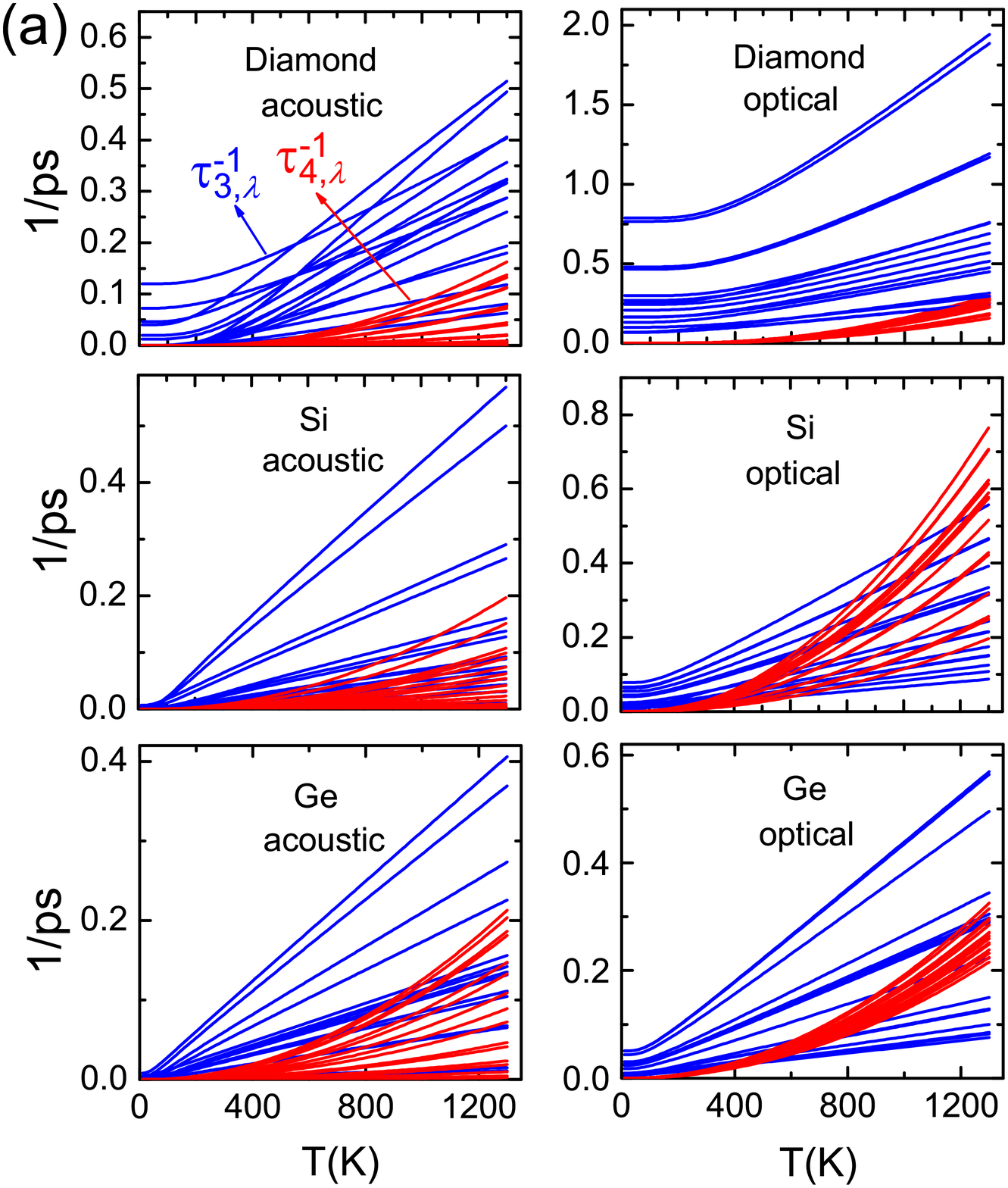}
	\includegraphics[width= 0.45\linewidth]{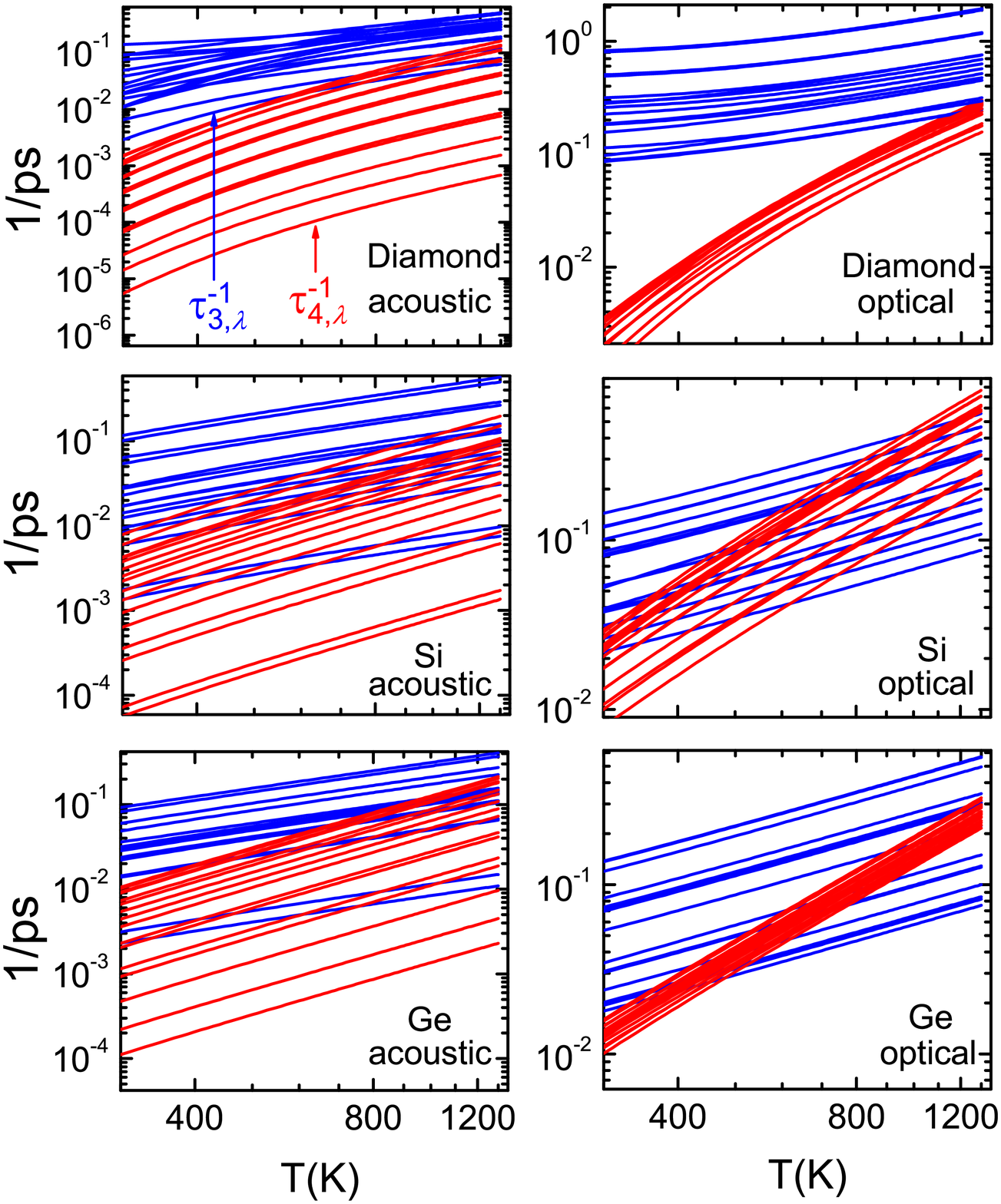}
	\caption{The $\tau_{3,\lambda}^{-1}$ (blue) and $\tau_{4,\lambda}^{-1}$ (red) of all the resolvable modes (excluding the $\Gamma$ point) from $\Gamma$ to $X$ as a function of temperature in diamond, Si and Ge. Each subfigure contains 32 curves. The 16 blue curves are $\tau_{3,\lambda}^{-1}$ for eight longitudinal and eight transverse modes with the reduced wave vectors of $\mathbf{k}^*$=($zeta$/8,0,0), where $zeta$ is an integer from 1 to 8. The 16 red curves are $\tau_{4,\lambda}^{-1}$ for these same modes. In (b), we show the logarithmic plots corresponding to plot (a) to demonstrate the power-law dependence of $\tau_{4,\lambda}^{-1}$ on temperature.}
	\label{t_all}
\end{figure}

\begin{figure}[tbph]
	\centering
	\includegraphics[width= 3.5in]{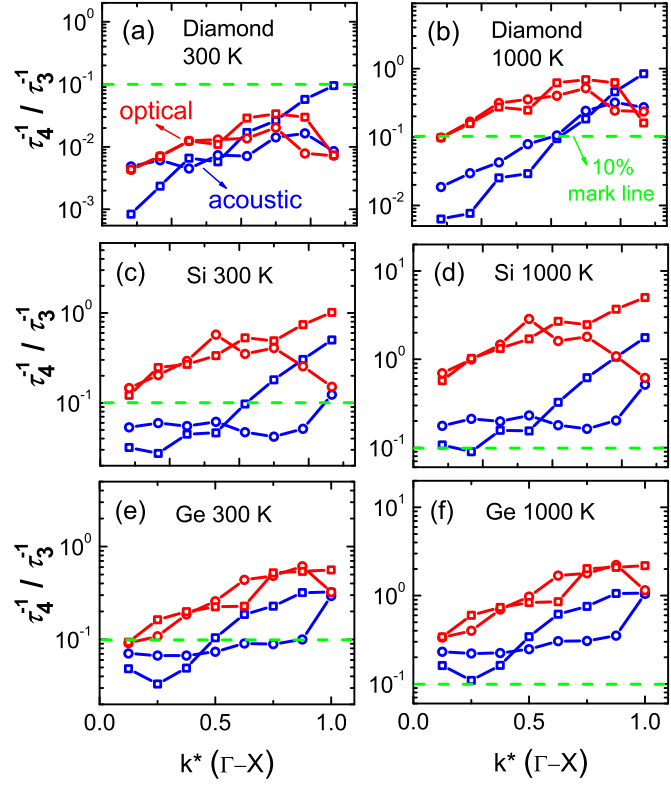}
	\caption{The ratio $\tau_{4,\lambda}^{-1}/\tau_{3,\lambda}^{-1}$ with respect to the reduced wave vector ($\Gamma$-X) for the TA [blue square], LA [blue circle], TO [red square] and LO [red circle] branches at 300 K and 1000 K in (a,b) diamond, (c,d) silicon and (e,f) germanium. The green dashed lines at $\tau_{4,\lambda}^{-1}/\tau_{3,\lambda}^{-1}$=10\% help to guide the eye.}\label{fig_t43CSiGe}
\end{figure}

\begin{figure}[tbph]
	\begin{center}
		\includegraphics[width= 0.46\textwidth]{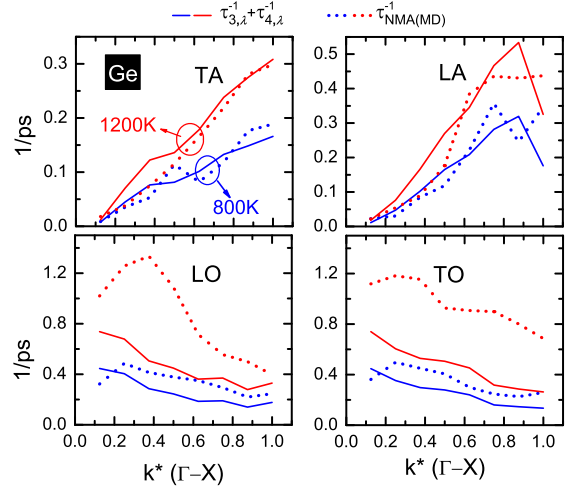}
	\end{center}
	\caption{The comparison between $\tau_{3,\lambda}^{-1}+\tau_{4,\lambda}^{-1}$ (solid curves) and $\tau_{{\rm NMA},\lambda}^{-1}$ (dashed curves) in Ge as a function of the reduced wave vector ($\Gamma$-$X$) at 800 K (blue) and 1200 K (red), for the TA, LA, LO, and TO branches.}\label{tGe}
\end{figure}

The importance of four-phonon scattering in less anharmonic materials -- diamond, silicon and germanium is studied. As shown in Fig.\,\ref{t_all}, the general temperature dependencies of $\tau_{3,\lambda}^{-1}\sim T$ and $\tau_{4,\lambda}^{-1}\sim T^2$ have been observed for both acoustic and optical phonons in all these materials. In Fig.\,\ref{fig_t43CSiGe}, we choose $T=$ 300 and 1000 K to show the relative importance of four-phonon scattering, $\tau_{4,\lambda}^{-1}/\tau_{3,\lambda}^{-1}$, as a function of the reduced wave vector from $\Gamma$ to $X$. At room temperature, $\tau_{4,\lambda}^{-1}/\tau_{3,\lambda}^{-1}$ for acoustic branches is roughly below 0.1, confirming less anharmonicities in these three materials than in argon. As $T$ rises to 1000 K, $\tau_{4,\lambda}^{-1}/\tau_{3,\lambda}^{-1}$ increases to 0.1-1 for most acoustic phonons in silicon and germanium, indicating that four-phonon scattering becomes comparable to three-phonon scattering. In comparison, with the same lattice structure, diamond has the strongest bonding strength and the least anharmonicity while germanium has the softest bonds and the most anharmonicity. In Figs.\,\ref{t_all} and\,\ref{fig_t43CSiGe}, it is clearly seen that four-phonon scattering is more important for more strongly anharmonic materials and higher temperatures. In contrast to acoustic phonons, optical phonons typically have much higher four-phonon scattering rates which are comparable to three-phonon scattering rates even at low temperatures. The accuracy of the results has been demonstrated by the general agreement between $\tau_{3,\lambda}^{-1}+\tau_{4,\lambda}^{-1}$ and $\tau_{{\rm NMA},\lambda}^{-1}$. In Fig.\,\ref{tGe}, we compare $\tau_{3,\lambda}^{-1}+\tau_{4,\lambda}^{-1}$ to $\tau_{{\rm NMA},\lambda}^{-1}$ in Ge at high temperatures of 800 K and 1200 K. Reasonable agreement is found for the acoustic phonons considering the uncertainty of MD simulations. If four-phonon scattering is excluded, no agreement can be achieved. One interesting finding is that $\tau_{3,\lambda}^{-1}+\tau_{4,\lambda}^{-1}$ of optical phonons is typically lower than $\tau_{{\rm NMA},\lambda}^{-1}$, indicating a possibility of high five-phonon scattering rates of optical phonons.

\begin{figure}[tpbh]
	\centering
	\includegraphics[width= 3.5in]{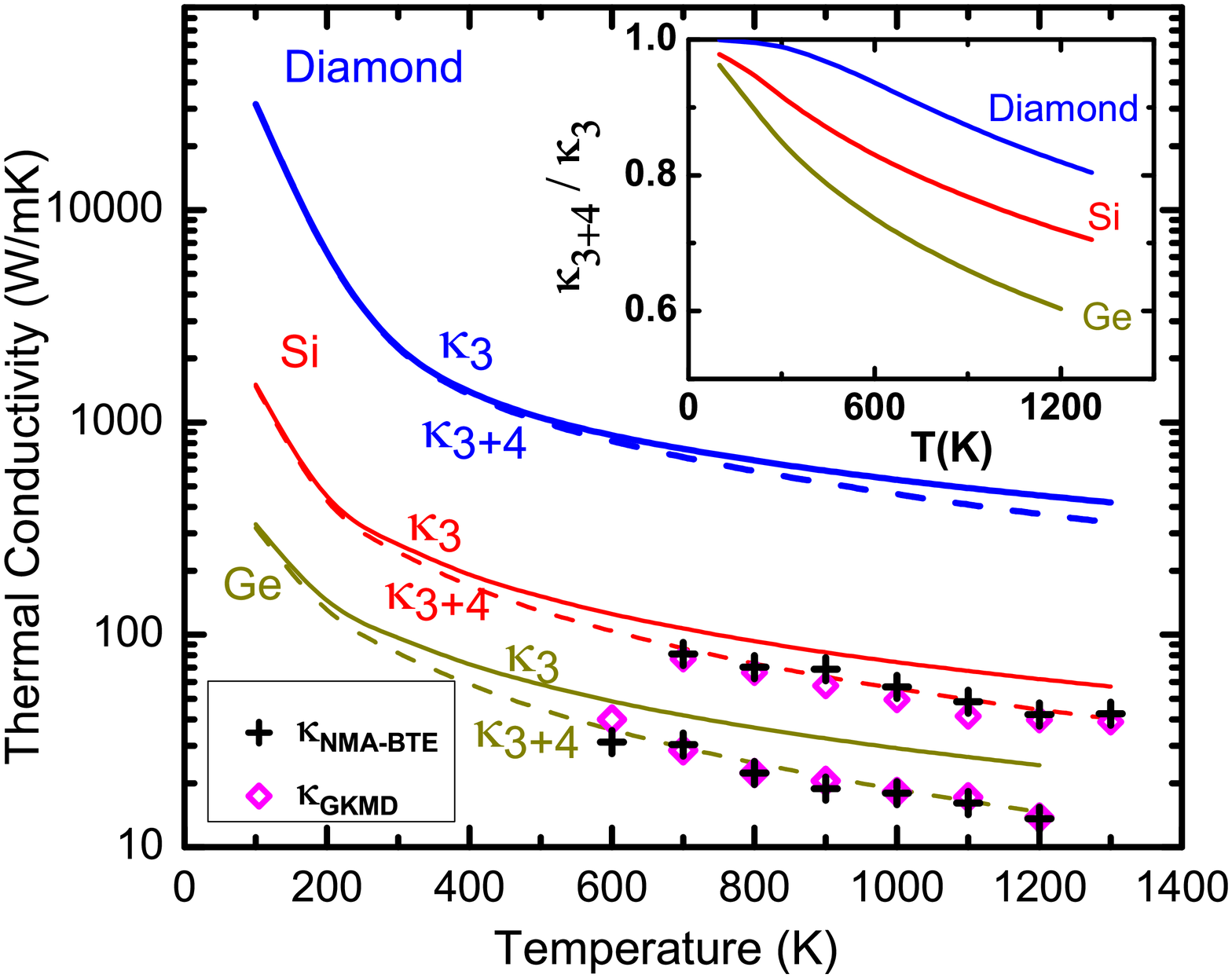}
	\caption{The lattice thermal conductivity $\kappa$ values of diamond, silicon, and germanium predicted from $\tau_{3,\lambda}^{-1}$, $\tau_{3,\lambda}^{-1}+\tau_{4,\lambda}^{-1}$, $\tau_{{\rm NMA},\lambda}^{-1}$, and the GK method as a function of temperature, with the inset showing the ratio $\kappa_{3+4}/\kappa_{3}$.}\label{fig_CSiGe}
\end{figure}

The lattice thermal conductivities of diamond, silicon and germanium are shown in Fig.\,\ref{fig_CSiGe}. $\kappa_3$ and $\kappa_{3+4}$ match well with each other at low temperatures, indicating that four-phonon scattering is negligible. At room temperature, $\kappa_{3+4}$ is lower than $\kappa_3$ by 1\%, 8\%, and 15\% for diamond, silicon, and germanium, respectively, as shown in the inset. As the temperature increases to 1000 K, such discrepancy grows to 15\%, 25\%, and 36\%, respectively. The discrepancy between a previously predicted $\kappa_3$ of Si at 1000 K using first principles\,\cite{Esf2011prb} and the experimental value\,\cite{Glass1964PR} is about 27\%, which is consistent with our calculations. Such results indicate that even in weakly anharmonic materials, four-phonon scattering may play a critical role at high temperatures. A good agreement between $\kappa_{3+4}$ and $\kappa_{\rm NMA}$ as well as $\kappa_{\rm GK(MD)}$ is found for silicon and germanium at high temperatures in Fig.\,\ref{fig_CSiGe}. The comparison in diamond is not done since diamond has a high Debye temperature ($\sim$2200 K), below which $\kappa_{3+4}$ obtained from quantum mechanics is not comparable to $\kappa_{\rm NMA}$ and $\kappa_{\rm GK(MD)}$ from classical MD. Since we use empirical interatomic potentials which are approximations to the true atomic interactions, the numbers presented here should be understood with caution or on a semi-quantitative basis.

\section{Discussion}

\subsection{Issue in the Boltzmann distribution-based ALD formula}
\label{Sec_Boltz}

In Sec.\,\ref{Sec_formula}, the ALD formulas are derived by using Bose-Einstein distribution starting from Eq.\,(\ref{eq_math_0}). In the following part, we derive the ALD formula based on Boltzmann distribution, taking three-phonon scattering as an example. Equations\,(\ref{eq_math_0}) and (\ref{eq_math1}), or the relation $\lambda\!\rightarrow\!\lambda_1\!+\!\lambda_2$: $1+\frac{1}{n_\lambda^0}=(1+\frac{1}{n_{\lambda_1}^0})(1+\frac{1}{n_{\lambda_2}^0})$ become
\begin{equation}
	\label{eq_Boltzmann1}
	\lambda\!\rightarrow\!\lambda_1\!+\!\lambda_2\!: \frac{1}{n_\lambda^0}=\frac{1}{n_{\lambda_1}^0}+\frac{1}{n_{\lambda_2}^0}.
\end{equation}
Equations\,(\ref{eq_math_2}) and (\ref{eq_math2}), or the relation $\lambda\!+\lambda_1\!\rightarrow\!\lambda_2$: $(1+\frac{1}{n_\lambda^0})(1+\frac{1}{n_{\lambda_1}^0})=1+\frac{1}{n_{\lambda_2}^0}$ become
\begin{equation}
	\label{eq_Boltzmann2}
	\lambda\!+\lambda_1\!\rightarrow\!\lambda_2\!: \frac{1}{n_\lambda^0}+\frac{1}{n_{\lambda_1}^0}=\frac{1}{n_{\lambda_2}^0}.
\end{equation}
By substituting Eqs.\,(\ref{eq_n})-(\ref{eq_n3prime}) into Eq.\,(\ref{BTE2}) by using the relations in Eqs.\,(\ref{eq_Boltzmann1}) and (\ref{eq_Boltzmann2}), we obtain
\begin{equation}
	\label{BTE3_B}
	\frac{\partial n_\lambda'}{\partial t} |_s = -\sum_{\lambda_1\lambda_2}\left\{   \frac{1}{2}\left[(1+n_{\lambda_1}^0+n_{\lambda_2}^0)n_\lambda'+n_\lambda^0\right]\mathcal{L}_- \!+\! \left[(n_{\lambda_1}^0- n_{\lambda_2}^0)n_\lambda'-n_{\lambda_2}^0\right]\mathcal{L}_+ \right\}.
\end{equation}
This equation is found to contain two additional constant terms $+n_\lambda^0$ and $-n_{\lambda_2}^0$ in the brackets, compared to Eq.\,(\ref{BTE3}) based on the Bose-Einstein distribution. The constant terms lead the decay of the perturbation $n_\lambda'$ to be not exponential, and thus the exact relaxation time cannot be well defined, unless the two terms can be neglected or can cancel off with each other during the summation over $\lambda_1,\lambda_2$. However, we have found that they are not negligible. Also, the cancellation is not guaranteed. For example, if $\lambda$ is an optical phonon near $\Gamma$ point, the right-hand side of Eq.\,(\ref{BTE3_B}) only contains the first half (the splitting process).

Therefore, it is not an exact approach to directly employ Boltzmann occupation number in the Bose-Einstein distribution-based ALD formula to capture the classical effect. The Boltzmann distribution-based ALD formula cannot well define the phonon relaxation time.

\subsection{Role of normal process} 
\begin{figure}[tphb]
	\centering
	\includegraphics[width= 3.3in]{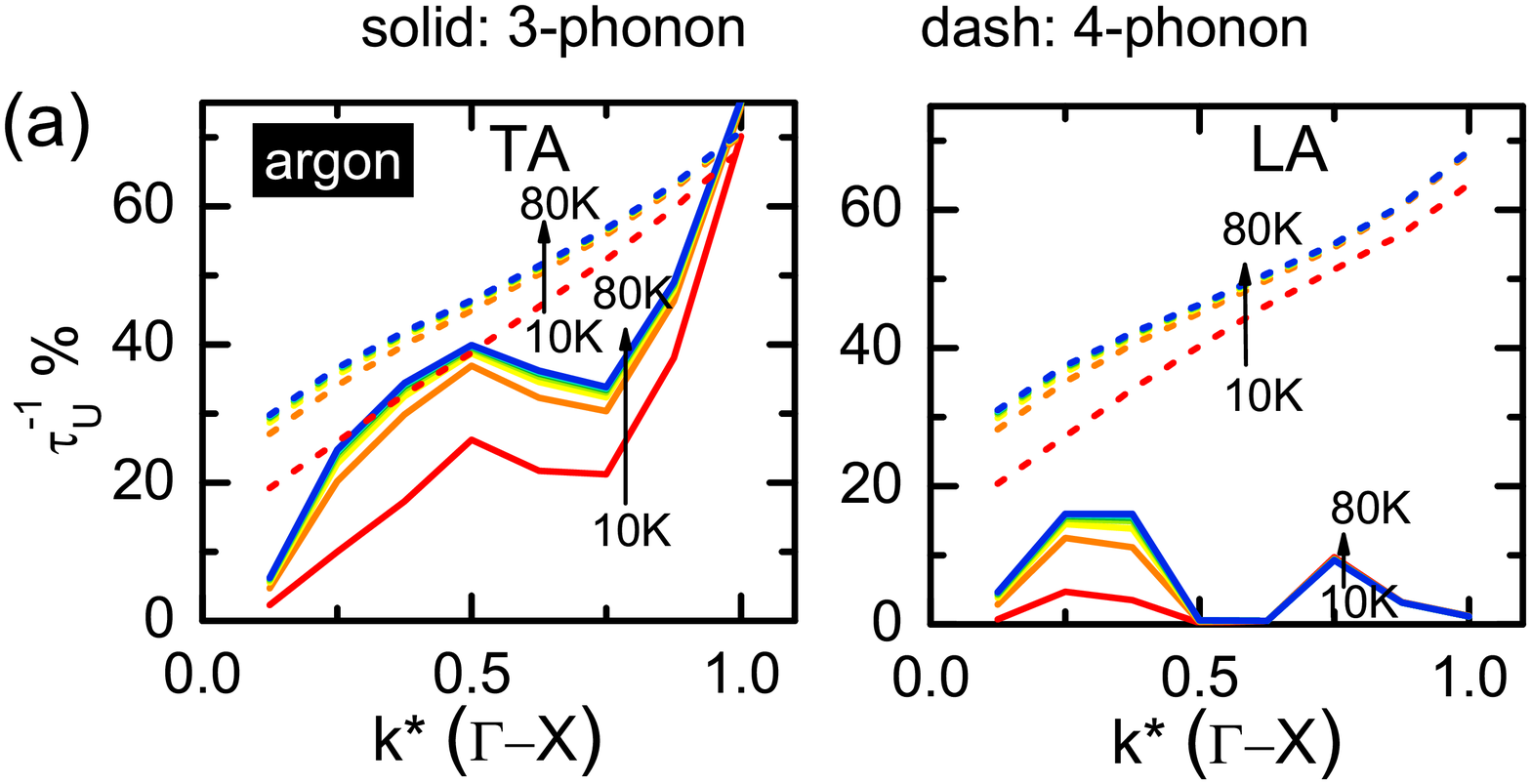}
	\includegraphics[width= 3.3in]{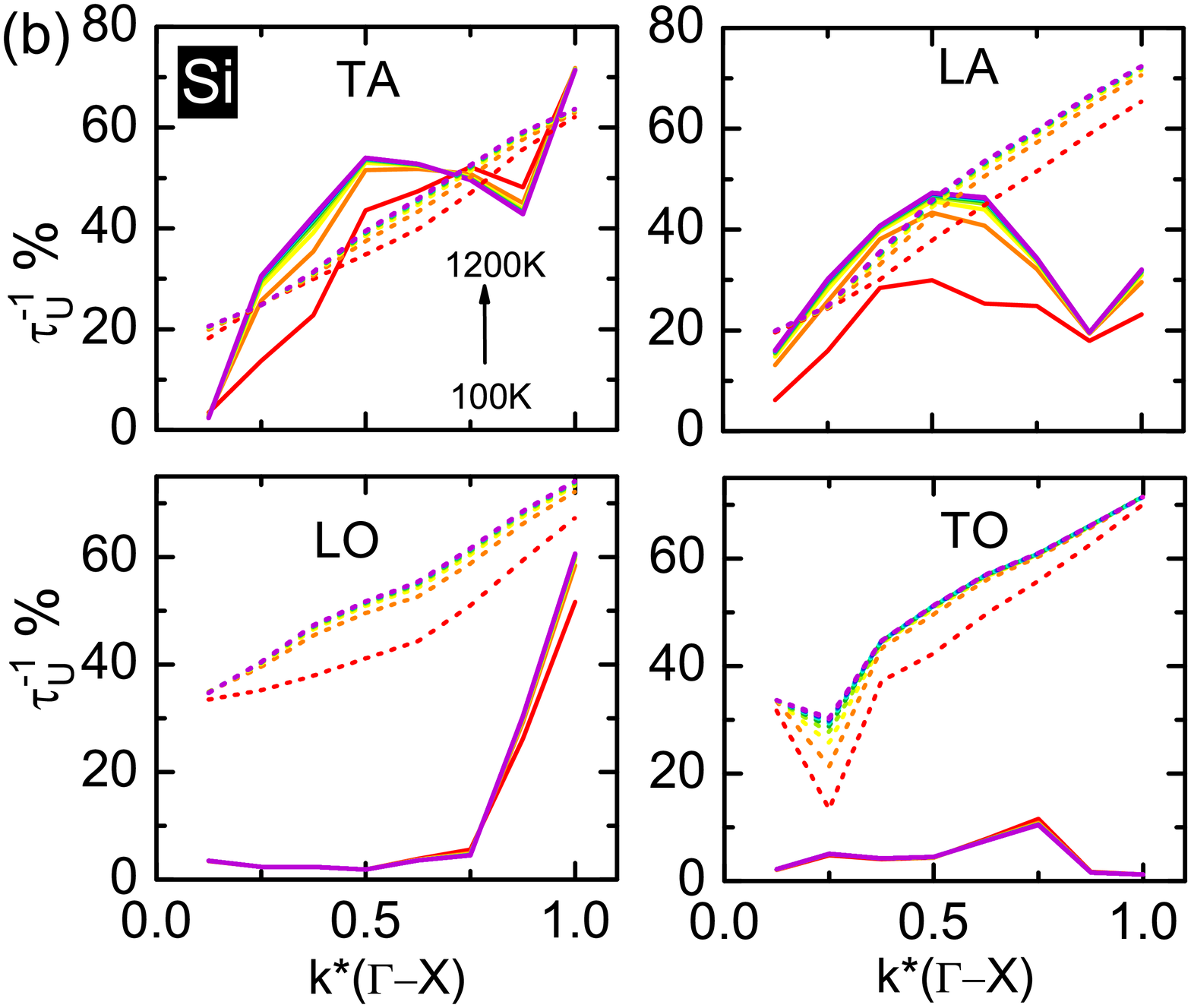}
	\caption{(a) Ar and (b) Si are taken as examples to show the percentage of $U$ processes in three-phonon (solid curves) and four-phonon (dashed curves) scattering. The wavevector is along $\Gamma$-$X$. The temperature is 10-80 K for argon, and 100-1200 K for Si. The temperatures from low to high are represented by the colors from red to purple, and the corresponding curves are from bottom to top.}\label{U_all}
\end{figure}

For some materials an appropriate handling of $U$ and $N$ processes is important for the prediction of $\kappa$\,\cite{Omini1995pb}. For example, when $N$ processes dominate, the scattering does not introduce thermal resistance directly, and an exact solution to the linearized BTE is required beyond the SMRTA\,\cite{Omini1995pb, Broido2007apl}. Such physics has been found to be important in three-phonon scattering in graphene\,\cite{Lindsay2010SLG} where $\tau_{U}^{-1}\%$ is very low. In this work, however, the SMRTA is still valid for the phonon transport in argon, silicon, and germanium since $\tau_{U}^{-1}\%$, especially for acoustic phonons which dominate lattice thermal conductivity, is not low for either three-phonon scattering\,\cite{Ward2010prb} or four-phonon scattering. The latter is found to have a $\tau_{U}^{-1}\%$ that increases monotonically with increasing temperature and wave vector, as shown in Fig.\,\ref{U_all}, where we show the percentage $\tau_{U}^{-1}$\% of the Umklapp scattering rates of the total scattering rates for three-phonon processes, $\tau_{3,\lambda,U}^{-1}/\tau_{3,\lambda}^{-1}$, and four-phonon processes, $\tau_{4,\lambda,U}^{-1}/\tau_{4,\lambda}^{-1}$. For both acoustic and optical phonons, $\tau_U^{-1}$\% in three- and four-phonon scattering increases monotonically with increasing temperature. For acoustic phonons, three and four-phonon processes have similar $\tau_U^{-1}$\%; whereas for optical branches, four-phonon scattering has much higher $\tau_U^{-1}$\% than three-phonon scattering.

\subsection{Negligible three-phonon to the second order}

\begin{figure}[htpb]
	\begin{center}
		\includegraphics[width= 0.5\linewidth]{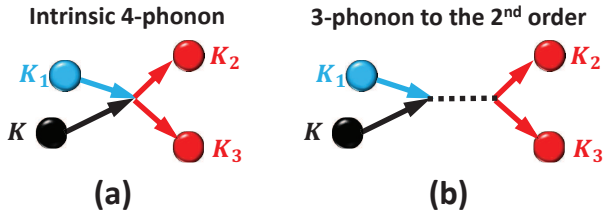}
	\end{center}
	\caption{The diagram examples for the comparison between (a) the intrinsic four-phonon scattering and (b) the three-phonon scattering to the second order.}\label{fig_3_2nd}
\end{figure}

We note that two three-phonon processes: $\lambda_1+\lambda_2\rightarrow\lambda'$ and $\lambda'\rightarrow\lambda_3+\lambda_4$ may be combined to give the three-phonon scattering to the second order, which is another type of fourth-order process\,\cite{Ziman_book}, as shown in Fig.\,\ref{fig_3_2nd} (b). Here $\lambda'$ is an intermediate virtual state. The energy is conserved from the initial state $\lambda_1+\lambda_2$ to the final state $\lambda_3+\lambda_4$, while the energy is not necessarily conserved in the first step or in the second step alone\,\cite{Ziman_book}. The energy denominators of 3-phonon scattering
\begin{equation}
	\frac{\langle i|\hat{H}_3|f\rangle}{|E_i-E_f|}
	\label{eq_H3_transition}
\end{equation}
and four-phonon scattering
\begin{equation}
	\frac{\langle i|\hat{H}_4|f\rangle}{|E_i-E_f|}
	\label{eq_H4_transition}
\end{equation}
vanish due to the energy conservation law  $E_i=E_f$. Here $|i\rangle$ and $|f\rangle$ represent the initial and final states respectively. For example in three-phonon scattering $\lambda_1+\lambda_2\rightarrow\lambda_3$, $|i\rangle$ represents the state $|n_{\lambda_1}+1,n_{\lambda_2}+1,n_{\lambda_3}\rangle$, and $|f\rangle$ represents the state $|n_{\lambda_1},n_{\lambda_2},n_{\lambda_3}+1\rangle$. In contrast to Eqs.\,(\ref{eq_H3_transition}) and (\ref{eq_H4_transition}), the transition matrix element in the combined three-phonon process is
\begin{equation}
	\frac{\langle i|\hat{H}_3|vir\rangle\langle vir|\hat{H}_3|f\rangle}{|E_i-E_{vir}|}.
	\label{eq_H32_transition}
\end{equation}
The discussion of the denominator in Eq.\,(\ref{eq_H32_transition}) can be divided into two cases. In Case 1, the energy is not conserved in the first or the second step\,\cite{Ziman_book}. The energy denominators for the transition are not small. Therefore, the transition rate is considered to be not large as discussed in Ref.\,\cite{Carruthers1962pr1}. In Case 2, the energy conservation condition for the first step is nearly satisfied or satisfied. This process was named as ``the resonance in three-phonon scattering'' and discussed by Carruthers\,\cite{Carruthers1962pr1}. In this case, although the scattering is in the same order with the intrinsic four-phonon scattering, the number of scattering events that satisfy the energy and momentum selection rule is only $10^{-3}-10^{-5}$ of that in the intrinsic four-phonon scattering in our study. This is because the resonant three-phonon scattering has a strong requirement that the intermediate state has to be an existing phonon mode in the $\mathbf{k}$-mesh, while the intrinsic four-phonon scattering has no such requirement. For example, for Si with a $16\times16\times16$ $\mathbf{k}$-mesh and the energy conservation tolerant range as 1.24 meV (0.3 THz), the TA mode at $\mathbf{k}^*=(0.5,0,0)$ has 4.6$\times10^{7}$ intrinsic four-phonon events, and only 2.7$\times10^4$ resonant three-phonon events. For the TA mode at $\mathbf{k}^*=(0.625,0,0)$, the number of intrinsic four-phonon events is similarly about 4.6$\times10^{7}$ while the number of resonant three-phonon events is only 36. Therefore, the overall three phonon to the second order scattering rate is negligible compared to the intrinsic four-phonon scattering and thus is not considered in our work. Nevertheless, it is definitely worth a quantitative study in the future.

\section{Conclusions}
To conclude, a rigorous and direct method to calculate four-phonon scattering rates $\tau_{4,\lambda}^{-1}$ within the ALD framework has been developed. We have obtained $\tau_{4,\lambda}^{-1}$ by explicitly determining quantum mechanical scattering probability matrices for the full Brillouin zone. By investigating the bulk argon, diamond, silicon and germanium, we have found the key features of four-phonon scattering: (1) $\tau_{4,\lambda}^{-1}$ increases quadratically with temperature, one order higher than $\tau_{3,\lambda}^{-1}$, (2) $\tau_{4,\lambda}^{-1}$ is more important in more strongly anharmonic bulk materials, (3) for optical phonons, the fourth and higher order phonon scattering is much more important than three-phonon scattering even at low temperature, such finding could be also important in the studies of optical properties, electron-phonon coupling, photovoltaics, etc\,\cite{Bao2012jqsrt}, (4) the relative ratio of Umklapp scattering rate in four-phonon process is generally comparable or even larger than that in three-phonon process, and (5) the three-phonon to the second order is negligible compared to four-phonon process, although they are in the same perturbation order. Particularly, $\tau_{4,\lambda}^{-1}$ can reduce the thermal conductivity of Si by $\sim$25\% at 1000 K. Existing practice of ALD is limited to three-phonon scattering so the accuracy is only guaranteed for relatively low temperature. Now by including $\tau_{4,\lambda}^{-1}$ with our approach, ALD will be applicable for both low and high temperatures.

\section*{ACKNOWLEDGMENTS}
	We appreciate the help from Alan J. H. McGaughey, Lucas Lindsay, and Christopher A Robinson for proofreading our manuscript. Simulations were preformed at the Rosen Center for Advanced Computing (RCAC) of Purdue University. The work was partially supported by the National Science Foundation (Award No. 1150948) and the Defense Advanced Research Projects Agency (Award No. HR0011-15-2-0037).

\appendix

\section{Normal mode analysis and Green-Kubo method based on molecular dynamics}
\label{append}

The linewidth $\tau_{NMA,\lambda}^{-1}$ is obtained by performing the following NMA\,\cite{Dove_book,Qiu2011arXiv, Feng2014Jn,Feng2015jap} based on MD simulations,
\begin{eqnarray}
\dot{q}_\lambda(t) &=& \sum_\alpha^3\sum_b^{n_b}\sum_l^{N_c}\sqrt{\frac{m_b}{N_c}}\dot{u}_\alpha^{l,b}(t)e_{b,\alpha}^{\lambda *}\exp\left[i\mathbf{k}\cdot \mathbf{r}_0^l\right], \label{normal-mode} \\
E_\lambda(\omega) &=& \left| \mathcal{F} [\dot{q}_\lambda(t)] \right|^2= \frac{C_\lambda}{(\omega-\omega_\lambda^A)^2+(\tau_{NMA,\lambda}^{-1})^2/4}. \label{SEDfunction} 
\end{eqnarray}
Here, $\dot{u}_\alpha^{l,b}(t)$ is the $\alpha$th component of the time dependent velocity of the $b$th basis atom in the $l$th unit cell, $e$ is the phonon eigenvector, and $\mathbf{r}_0$ is the equilibrium position. $\mathcal{F}$ denotes the Fourier transformation. The spectral energy density $E_\lambda(\omega)$ of the phonon mode $\lambda$ is obtained by substituting $\dot{u}_\alpha^{l,b}(t)$ extracted from MD trajectory into Eq.\,(\ref{SEDfunction}), where $C_\lambda$ is a constant for a given $\lambda$. By fitting the spectral energy density as a Lorentzian function, the peak position $\omega_\lambda^A$ and full linewidth $\tau_{NMA,\lambda}^{-1}$ at half maximum are obtained. Our former work\,\cite{Feng2015jap} has shown that Eqs. (\ref{normal-mode}) and (\ref{SEDfunction}) are equivalent to another version of frequency-domain NMA that does not include phonon eigenvectors\,\cite{Koker2009prl,Thomas2010prb}
\begin{equation}
\label{Phi2}
\Phi(\mathbf{k},\omega)\! =\sum_j^{3n_b} E_\lambda(\omega)=\! \frac{1}{4\pi t_0}\sum_{\alpha}^{3}\sum_{b}^{n_b}\frac{m_b}{N_c}\left|\sum_l^{N_c}\! \int_0^{t_0}\! \dot{u}_\alpha^{l,b}(t) \exp(i\mathbf{k}\! \cdot\! \mathbf{r}_0^l \!-\! i\omega t)dt \right |^2.
\end{equation}
A full discussion about the methods of predicting phonon relaxation time was given in Ref.\,\cite{Feng2014Jn}. The simulation domains for those materials studied in our work are $8\times8\times8$ conventional cells (2048 atoms for Ar, and 4096 atoms for Si and Ge). Nine $\mathbf{k}$-points with the reduced wavevector being $\mathbf{k}^*=(\zeta/8,0,0)$ are resolved from $\Gamma$ to $X$ in the BZ, where $\zeta$ is an integer from 0 to 8.

In the Green-Kubo method\,\cite{GreenKubo}, the thermal conductivity is given by
\begin{equation}
\label{eq_GK}
\kappa_z=\frac{1}{k_BT^2V}\int_0^\infty\langle S_z(t)S_z(0)\rangle dt,
\end{equation}
where $z$ denotes the transport direction, $V$ is the simulation domain volume, and $S_z$ represents the heat current in the $z$ direction. The time step interval and total simulation time are set as 0.5 fs and 5 ns, respectively. The autocorrelation length is set as 500 ps, which is long enough to obtain converged heat current auto-correlation functions (HCACF) for argon, Si, and Ge since most phonon relaxation times are far below 500 ps at high temperatures. The NMA and the GK-MD method are both based on equilibrium MD which has much less size effect than non-equilibrium MD. Our GK-MD simulations show that typically $8\times8\times8$ cells are large enough to get a converged thermal conductivity of those bulk materials at the temperatures studied.

\bibliography{../../../bibfile}

\begin{thebibliography}{35}%
\makeatletter
\providecommand \@ifxundefined [1]{%
 \@ifx{#1\undefined}
}%
\providecommand \@ifnum [1]{%
 \ifnum #1\expandafter \@firstoftwo
 \else \expandafter \@secondoftwo
 \fi
}%
\providecommand \@ifx [1]{%
 \ifx #1\expandafter \@firstoftwo
 \else \expandafter \@secondoftwo
 \fi
}%
\providecommand \natexlab [1]{#1}%
\providecommand \enquote  [1]{``#1''}%
\providecommand \bibnamefont  [1]{#1}%
\providecommand \bibfnamefont [1]{#1}%
\providecommand \citenamefont [1]{#1}%
\providecommand \href@noop [0]{\@secondoftwo}%
\providecommand \href [0]{\begingroup \@sanitize@url \@href}%
\providecommand \@href[1]{\@@startlink{#1}\@@href}%
\providecommand \@@href[1]{\endgroup#1\@@endlink}%
\providecommand \@sanitize@url [0]{\catcode `\\12\catcode `\$12\catcode
  `\&12\catcode `\#12\catcode `\^12\catcode `\_12\catcode `\%12\relax}%
\providecommand \@@startlink[1]{}%
\providecommand \@@endlink[0]{}%
\providecommand \url  [0]{\begingroup\@sanitize@url \@url }%
\providecommand \@url [1]{\endgroup\@href {#1}{\urlprefix }}%
\providecommand \urlprefix  [0]{URL }%
\providecommand \Eprint [0]{\href }%
\providecommand \doibase [0]{http://dx.doi.org/}%
\providecommand \selectlanguage [0]{\@gobble}%
\providecommand \bibinfo  [0]{\@secondoftwo}%
\providecommand \bibfield  [0]{\@secondoftwo}%
\providecommand \translation [1]{[#1]}%
\providecommand \BibitemOpen [0]{}%
\providecommand \bibitemStop [0]{}%
\providecommand \bibitemNoStop [0]{.\EOS\space}%
\providecommand \EOS [0]{\spacefactor3000\relax}%
\providecommand \BibitemShut  [1]{\csname bibitem#1\endcsname}%
\let\auto@bib@innerbib\@empty
\bibitem [{\citenamefont {Srivastava}(1990)}]{Srivastava_book}%
  \BibitemOpen
  \bibfield  {author} {\bibinfo {author} {\bibfnamefont {G.~P.}\ \bibnamefont
  {Srivastava}},\ }\href@noop {} {\emph {\bibinfo {title} {{The physics of
  phonons}}}}\ (\bibinfo  {publisher} {CRC Press},\ \bibinfo {address} {Adam
  Hilger, Bristol},\ \bibinfo {year} {1990})\BibitemShut {NoStop}%
\bibitem [{\citenamefont {Broido}\ \emph {et~al.}(2007)\citenamefont {Broido},
  \citenamefont {Malorny}, \citenamefont {Birner}, \citenamefont {Mingo},\ and\
  \citenamefont {Stewart}}]{Broido2007apl}%
  \BibitemOpen
  \bibfield  {author} {\bibinfo {author} {\bibfnamefont {D.~A.}\ \bibnamefont
  {Broido}}, \bibinfo {author} {\bibfnamefont {M.}~\bibnamefont {Malorny}},
  \bibinfo {author} {\bibfnamefont {G.}~\bibnamefont {Birner}}, \bibinfo
  {author} {\bibfnamefont {N.}~\bibnamefont {Mingo}}, \ and\ \bibinfo {author}
  {\bibfnamefont {D.~A.}\ \bibnamefont {Stewart}},\ }\href {\doibase
  10.1063/1.2822891} {\bibfield  {journal} {\bibinfo  {journal} {Applied
  Physics Letters}\ }\textbf {\bibinfo {volume} {91}},\ \bibinfo {pages}
  {231922} (\bibinfo {year} {2007})}\BibitemShut {NoStop}%
\bibitem [{\citenamefont {Maradudin}\ and\ \citenamefont
  {Fein}(1962)}]{Maradudin1962prb}%
  \BibitemOpen
  \bibfield  {author} {\bibinfo {author} {\bibfnamefont {A.}~\bibnamefont
  {Maradudin}}\ and\ \bibinfo {author} {\bibfnamefont {A.}~\bibnamefont
  {Fein}},\ }\href {http://prola.aps.org/abstract/PR/v128/i6/p2589\_1}
  {\bibfield  {journal} {\bibinfo  {journal} {Physical Review}\ }\textbf
  {\bibinfo {volume} {128}},\ \bibinfo {pages} {2589} (\bibinfo {year}
  {1962})}\BibitemShut {NoStop}%
\bibitem [{\citenamefont {Maradudin}\ \emph {et~al.}(1962)\citenamefont
  {Maradudin}, \citenamefont {Fein},\ and\ \citenamefont
  {Vineyard}}]{Maradudin1962pss}%
  \BibitemOpen
  \bibfield  {author} {\bibinfo {author} {\bibfnamefont {A.~A.}\ \bibnamefont
  {Maradudin}}, \bibinfo {author} {\bibfnamefont {A.~E.}\ \bibnamefont {Fein}},
  \ and\ \bibinfo {author} {\bibfnamefont {G.~H.}\ \bibnamefont {Vineyard}},\
  }\href {\doibase 10.1002/pssb.19620021106} {\bibfield  {journal} {\bibinfo
  {journal} {Physica Status Solidi (B)}\ }\textbf {\bibinfo {volume} {2}},\
  \bibinfo {pages} {1479} (\bibinfo {year} {1962})}\BibitemShut {NoStop}%
\bibitem [{\citenamefont {Debernardi}\ \emph {et~al.}(1995)\citenamefont
  {Debernardi}, \citenamefont {Baroni},\ and\ \citenamefont
  {Molinari}}]{Deb1995prl}%
  \BibitemOpen
  \bibfield  {author} {\bibinfo {author} {\bibfnamefont {A.}~\bibnamefont
  {Debernardi}}, \bibinfo {author} {\bibfnamefont {S.}~\bibnamefont {Baroni}},
  \ and\ \bibinfo {author} {\bibfnamefont {E.}~\bibnamefont {Molinari}},\
  }\href {\doibase 10.1103/PhysRevLett.75.1819} {\bibfield  {journal} {\bibinfo
   {journal} {Phys. Rev. Lett.}\ }\textbf {\bibinfo {volume} {75}},\ \bibinfo
  {pages} {1819} (\bibinfo {year} {1995})}\BibitemShut {NoStop}%
\bibitem [{\citenamefont {Turney}\ \emph {et~al.}(2009)\citenamefont {Turney},
  \citenamefont {Landry}, \citenamefont {McGaughey},\ and\ \citenamefont
  {Amon}}]{Turney2009prb1}%
  \BibitemOpen
  \bibfield  {author} {\bibinfo {author} {\bibfnamefont {J.}~\bibnamefont
  {Turney}}, \bibinfo {author} {\bibfnamefont {E.}~\bibnamefont {Landry}},
  \bibinfo {author} {\bibfnamefont {A.}~\bibnamefont {McGaughey}}, \ and\
  \bibinfo {author} {\bibfnamefont {C.}~\bibnamefont {Amon}},\ }\href {\doibase
  10.1103/PhysRevB.79.064301} {\bibfield  {journal} {\bibinfo  {journal}
  {Physical Review B}\ }\textbf {\bibinfo {volume} {79}},\ \bibinfo {pages}
  {064301} (\bibinfo {year} {2009})}\BibitemShut {NoStop}%
\bibitem [{\citenamefont {Lindsay}\ \emph {et~al.}(2009)\citenamefont
  {Lindsay}, \citenamefont {Broido},\ and\ \citenamefont
  {Mingo}}]{Lindsay2009prb}%
  \BibitemOpen
  \bibfield  {author} {\bibinfo {author} {\bibfnamefont {L.}~\bibnamefont
  {Lindsay}}, \bibinfo {author} {\bibfnamefont {D.}~\bibnamefont {Broido}}, \
  and\ \bibinfo {author} {\bibfnamefont {N.}~\bibnamefont {Mingo}},\ }\href
  {\doibase 10.1103/PhysRevB.80.125407} {\bibfield  {journal} {\bibinfo
  {journal} {Physical Review B}\ }\textbf {\bibinfo {volume} {80}},\ \bibinfo
  {pages} {125407} (\bibinfo {year} {2009})}\BibitemShut {NoStop}%
\bibitem [{\citenamefont {Esfarjani}\ \emph {et~al.}(2011)\citenamefont
  {Esfarjani}, \citenamefont {Chen},\ and\ \citenamefont
  {Stokes}}]{Esf2011prb}%
  \BibitemOpen
  \bibfield  {author} {\bibinfo {author} {\bibfnamefont {K.}~\bibnamefont
  {Esfarjani}}, \bibinfo {author} {\bibfnamefont {G.}~\bibnamefont {Chen}}, \
  and\ \bibinfo {author} {\bibfnamefont {H.~T.}\ \bibnamefont {Stokes}},\
  }\href {\doibase 10.1103/PhysRevB.84.085204} {\bibfield  {journal} {\bibinfo
  {journal} {Physical Review B}\ }\textbf {\bibinfo {volume} {84}},\ \bibinfo
  {pages} {085204} (\bibinfo {year} {2011})}\BibitemShut {NoStop}%
\bibitem [{\citenamefont {Lindsay}\ \emph {et~al.}(2012)\citenamefont
  {Lindsay}, \citenamefont {Broido},\ and\ \citenamefont
  {Reinecke}}]{Lindsay2012prl}%
  \BibitemOpen
  \bibfield  {author} {\bibinfo {author} {\bibfnamefont {L.}~\bibnamefont
  {Lindsay}}, \bibinfo {author} {\bibfnamefont {D.~a.}\ \bibnamefont {Broido}},
  \ and\ \bibinfo {author} {\bibfnamefont {T.~L.}\ \bibnamefont {Reinecke}},\
  }\href {\doibase 10.1103/PhysRevLett.109.095901} {\bibfield  {journal}
  {\bibinfo  {journal} {Physical Review Letters}\ }\textbf {\bibinfo {volume}
  {109}},\ \bibinfo {pages} {095901} (\bibinfo {year} {2012})}\BibitemShut
  {NoStop}%
\bibitem [{\citenamefont {Feng}\ and\ \citenamefont {Ruan}(2014)}]{Feng2014Jn}%
  \BibitemOpen
  \bibfield  {author} {\bibinfo {author} {\bibfnamefont {T.}~\bibnamefont
  {Feng}}\ and\ \bibinfo {author} {\bibfnamefont {X.}~\bibnamefont {Ruan}},\
  }\href {\doibase 10.1155/2014/206370} {\bibfield  {journal} {\bibinfo
  {journal} {Journal of Nanomaterials}\ }\textbf {\bibinfo {volume} {2014}},\
  \bibinfo {pages} {206370} (\bibinfo {year} {2014})}\BibitemShut {NoStop}%
\bibitem [{\citenamefont {Glassbrenner}\ and\ \citenamefont
  {Slack}(1964)}]{Glass1964PR}%
  \BibitemOpen
  \bibfield  {author} {\bibinfo {author} {\bibfnamefont {C.~J.}\ \bibnamefont
  {Glassbrenner}}\ and\ \bibinfo {author} {\bibfnamefont {G.~A.}\ \bibnamefont
  {Slack}},\ }\href {http://prola.aps.org/abstract/PR/v134/i4A/pA1058\_1}
  {\bibfield  {journal} {\bibinfo  {journal} {Physical Review}\ }\textbf
  {\bibinfo {volume} {134}},\ \bibinfo {pages} {A1058} (\bibinfo {year}
  {1964})}\BibitemShut {NoStop}%
\bibitem [{\citenamefont {Joshi}\ \emph {et~al.}(1970)\citenamefont {Joshi},
  \citenamefont {Tiwari},\ and\ \citenamefont {Verma}}]{Joshi1970prb}%
  \BibitemOpen
  \bibfield  {author} {\bibinfo {author} {\bibfnamefont {Y.~P.}\ \bibnamefont
  {Joshi}}, \bibinfo {author} {\bibfnamefont {M.~D.}\ \bibnamefont {Tiwari}}, \
  and\ \bibinfo {author} {\bibfnamefont {G.~S.}\ \bibnamefont {Verma}},\ }\href
  {\doibase 10.1103/PhysRevB.1.642} {\bibfield  {journal} {\bibinfo  {journal}
  {Phys. Rev. B}\ }\textbf {\bibinfo {volume} {1}},\ \bibinfo {pages} {642}
  (\bibinfo {year} {1970})}\BibitemShut {NoStop}%
\bibitem [{\citenamefont {Ecsedy}\ and\ \citenamefont
  {Klemens}(1977)}]{Ecsedy1977prb}%
  \BibitemOpen
  \bibfield  {author} {\bibinfo {author} {\bibfnamefont {D.~J.}\ \bibnamefont
  {Ecsedy}}\ and\ \bibinfo {author} {\bibfnamefont {P.~G.}\ \bibnamefont
  {Klemens}},\ }\href {\doibase 10.1103/PhysRevB.15.5957} {\bibfield  {journal}
  {\bibinfo  {journal} {Phys. Rev. B}\ }\textbf {\bibinfo {volume} {15}},\
  \bibinfo {pages} {5957} (\bibinfo {year} {1977})}\BibitemShut {NoStop}%
\bibitem [{\citenamefont {Lindsay}\ and\ \citenamefont
  {Broido}(2008)}]{Lindsay2008jp}%
  \BibitemOpen
  \bibfield  {author} {\bibinfo {author} {\bibfnamefont {L.}~\bibnamefont
  {Lindsay}}\ and\ \bibinfo {author} {\bibfnamefont {D.~a.}\ \bibnamefont
  {Broido}},\ }\href {\doibase 10.1088/0953-8984/20/16/165209} {\bibfield
  {journal} {\bibinfo  {journal} {Journal of Physics: Condensed Matter}\
  }\textbf {\bibinfo {volume} {20}},\ \bibinfo {pages} {165209} (\bibinfo
  {year} {2008})}\BibitemShut {NoStop}%
\bibitem [{\citenamefont {Sapna}\ and\ \citenamefont
  {Singh}(2013)}]{Sapna2013mplb}%
  \BibitemOpen
  \bibfield  {author} {\bibinfo {author} {\bibfnamefont {P.}~\bibnamefont
  {Sapna}}\ and\ \bibinfo {author} {\bibfnamefont {T.~J.}\ \bibnamefont
  {Singh}},\ }\href {\doibase 10.1142/S0217984913501170} {\bibfield  {journal}
  {\bibinfo  {journal} {Modern Physics Letters B}\ }\textbf {\bibinfo {volume}
  {27}},\ \bibinfo {pages} {1350117} (\bibinfo {year} {2013})}\BibitemShut
  {NoStop}%
\bibitem [{\citenamefont {Ziman}(1960)}]{Ziman_book}%
  \BibitemOpen
  \bibfield  {author} {\bibinfo {author} {\bibfnamefont {J.~M.}\ \bibnamefont
  {Ziman}},\ }\href@noop {} {\emph {\bibinfo {title} {{Electrons and
  Phonons}}}}\ (\bibinfo  {publisher} {Oxford University Press},\ \bibinfo
  {address} {London},\ \bibinfo {year} {1960})\BibitemShut {NoStop}%
\bibitem [{\citenamefont {Klemens}(1958)}]{Klemens_book}%
  \BibitemOpen
  \bibfield  {author} {\bibinfo {author} {\bibfnamefont {P.}~\bibnamefont
  {Klemens}},\ }\href@noop {} {\emph {\bibinfo {title} {{Solid State
  Physics}}}},\ Vol.~\bibinfo {volume} {7}\ (\bibinfo  {publisher} {Academic
  Press Inc.},\ \bibinfo {address} {New York, USA},\ \bibinfo {year}
  {1958})\BibitemShut {NoStop}%
\bibitem [{\citenamefont {Kaviany}(2008)}]{Kaviany_book}%
  \BibitemOpen
  \bibfield  {author} {\bibinfo {author} {\bibfnamefont {M.}~\bibnamefont
  {Kaviany}},\ }\href@noop {} {\emph {\bibinfo {title} {{Heat Transfer
  Physics}}}}\ (\bibinfo  {publisher} {Cambridge University Press},\ \bibinfo
  {address} {New York},\ \bibinfo {year} {2008})\BibitemShut {NoStop}%
\bibitem [{\citenamefont {Esfarjani}\ and\ \citenamefont
  {Stokes}(2008)}]{Esfar2008prb}%
  \BibitemOpen
  \bibfield  {author} {\bibinfo {author} {\bibfnamefont {K.}~\bibnamefont
  {Esfarjani}}\ and\ \bibinfo {author} {\bibfnamefont {H.~T.}\ \bibnamefont
  {Stokes}},\ }\href {\doibase 10.1103/PhysRevB.77.144112} {\bibfield
  {journal} {\bibinfo  {journal} {Phys. Rev. B}\ }\textbf {\bibinfo {volume}
  {77}},\ \bibinfo {pages} {144112} (\bibinfo {year} {2008})}\BibitemShut
  {NoStop}%
\bibitem [{\citenamefont {Li}\ \emph {et~al.}(2012)\citenamefont {Li},
  \citenamefont {Lindsay}, \citenamefont {Broido}, \citenamefont {Stewart},\
  and\ \citenamefont {Mingo}}]{Li2012prb}%
  \BibitemOpen
  \bibfield  {author} {\bibinfo {author} {\bibfnamefont {W.}~\bibnamefont
  {Li}}, \bibinfo {author} {\bibfnamefont {L.}~\bibnamefont {Lindsay}},
  \bibinfo {author} {\bibfnamefont {D.~A.}\ \bibnamefont {Broido}}, \bibinfo
  {author} {\bibfnamefont {D.~A.}\ \bibnamefont {Stewart}}, \ and\ \bibinfo
  {author} {\bibfnamefont {N.}~\bibnamefont {Mingo}},\ }\href {\doibase
  10.1103/PhysRevB.86.174307} {\bibfield  {journal} {\bibinfo  {journal} {Phys.
  Rev. B}\ }\textbf {\bibinfo {volume} {86}},\ \bibinfo {pages} {174307}
  (\bibinfo {year} {2012})}\BibitemShut {NoStop}%
\bibitem [{\citenamefont {Ashcroft}\ and\ \citenamefont
  {Mermin}(1976)}]{Ashcroft_book}%
  \BibitemOpen
  \bibfield  {author} {\bibinfo {author} {\bibfnamefont {N.~W.}\ \bibnamefont
  {Ashcroft}}\ and\ \bibinfo {author} {\bibfnamefont {N.~D.}\ \bibnamefont
  {Mermin}},\ }\href@noop {} {\emph {\bibinfo {title} {{Solid State
  Physics}}}}\ (\bibinfo  {publisher} {Saunders College Publishing},\ \bibinfo
  {address} {Fort Worth},\ \bibinfo {year} {1976})\BibitemShut {NoStop}%
\bibitem [{\citenamefont {Tersoff}(1989)}]{Tersoff1989}%
  \BibitemOpen
  \bibfield  {author} {\bibinfo {author} {\bibfnamefont {J.}~\bibnamefont
  {Tersoff}},\ }\href {\doibase 10.1103/PhysRevB.39.5566} {\bibfield  {journal}
  {\bibinfo  {journal} {Phys. Rev. B}\ }\textbf {\bibinfo {volume} {39}},\
  \bibinfo {pages} {5566} (\bibinfo {year} {1989})}\BibitemShut {NoStop}%
\bibitem [{\citenamefont {McGaughey}\ and\ \citenamefont
  {Kaviany}(2004)}]{McGaughey2004prb}%
  \BibitemOpen
  \bibfield  {author} {\bibinfo {author} {\bibfnamefont {A.}~\bibnamefont
  {McGaughey}}\ and\ \bibinfo {author} {\bibfnamefont {M.}~\bibnamefont
  {Kaviany}},\ }\href {\doibase 10.1103/PhysRevB.69.094303} {\bibfield
  {journal} {\bibinfo  {journal} {Physical Review B}\ }\textbf {\bibinfo
  {volume} {69}},\ \bibinfo {pages} {094303} (\bibinfo {year}
  {2004})}\BibitemShut {NoStop}%
\bibitem [{\citenamefont {Kaburaki}\ \emph {et~al.}(2007)\citenamefont
  {Kaburaki}, \citenamefont {Li}, \citenamefont {Yip},\ and\ \citenamefont
  {Kimizuka}}]{Kab2007jap}%
  \BibitemOpen
  \bibfield  {author} {\bibinfo {author} {\bibfnamefont {H.}~\bibnamefont
  {Kaburaki}}, \bibinfo {author} {\bibfnamefont {J.}~\bibnamefont {Li}},
  \bibinfo {author} {\bibfnamefont {S.}~\bibnamefont {Yip}}, \ and\ \bibinfo
  {author} {\bibfnamefont {H.}~\bibnamefont {Kimizuka}},\ }\href {\doibase
  10.1063/1.2772547} {\bibfield  {journal} {\bibinfo  {journal} {Journal of
  Applied Physics}\ }\textbf {\bibinfo {volume} {102}},\ \bibinfo {pages}
  {043514} (\bibinfo {year} {2007})}\BibitemShut {NoStop}%
\bibitem [{\citenamefont {Omini}\ and\ \citenamefont
  {Sparavigna}(1995)}]{Omini1995pb}%
  \BibitemOpen
  \bibfield  {author} {\bibinfo {author} {\bibfnamefont {M.}~\bibnamefont
  {Omini}}\ and\ \bibinfo {author} {\bibfnamefont {A.}~\bibnamefont
  {Sparavigna}},\ }\href {\doibase 10.1016/0921-4526(95)00016-3} {\bibfield
  {journal} {\bibinfo  {journal} {Physica B: Condensed Matter}\ }\textbf
  {\bibinfo {volume} {212}},\ \bibinfo {pages} {101} (\bibinfo {year}
  {1995})}\BibitemShut {NoStop}%
\bibitem [{\citenamefont {Lindsay}\ \emph {et~al.}(2010)\citenamefont
  {Lindsay}, \citenamefont {Broido},\ and\ \citenamefont
  {Mingo}}]{Lindsay2010SLG}%
  \BibitemOpen
  \bibfield  {author} {\bibinfo {author} {\bibfnamefont {L.}~\bibnamefont
  {Lindsay}}, \bibinfo {author} {\bibfnamefont {D.~A.}\ \bibnamefont {Broido}},
  \ and\ \bibinfo {author} {\bibfnamefont {N.}~\bibnamefont {Mingo}},\ }\href
  {\doibase 10.1103/PhysRevB.82.115427} {\bibfield  {journal} {\bibinfo
  {journal} {Physical Review B}\ }\textbf {\bibinfo {volume} {82}},\ \bibinfo
  {pages} {115427} (\bibinfo {year} {2010})}\BibitemShut {NoStop}%
\bibitem [{\citenamefont {Ward}\ and\ \citenamefont
  {Broido}(2010)}]{Ward2010prb}%
  \BibitemOpen
  \bibfield  {author} {\bibinfo {author} {\bibfnamefont {A.}~\bibnamefont
  {Ward}}\ and\ \bibinfo {author} {\bibfnamefont {D.~A.}\ \bibnamefont
  {Broido}},\ }\href {\doibase 10.1103/PhysRevB.81.085205} {\bibfield
  {journal} {\bibinfo  {journal} {Phys. Rev. B}\ }\textbf {\bibinfo {volume}
  {81}},\ \bibinfo {pages} {085205} (\bibinfo {year} {2010})}\BibitemShut
  {NoStop}%
\bibitem [{\citenamefont {Carruthers}(1962)}]{Carruthers1962pr1}%
  \BibitemOpen
  \bibfield  {author} {\bibinfo {author} {\bibfnamefont {P.}~\bibnamefont
  {Carruthers}},\ }\href {http://prola.aps.org/abstract/PR/v125/i1/p123\_1}
  {\bibfield  {journal} {\bibinfo  {journal} {Physical Review}\ }\textbf
  {\bibinfo {volume} {125}},\ \bibinfo {pages} {123} (\bibinfo {year}
  {1962})}\BibitemShut {NoStop}%
\bibitem [{\citenamefont {Bao}\ \emph {et~al.}(2012)\citenamefont {Bao},
  \citenamefont {Qiu}, \citenamefont {Zhang},\ and\ \citenamefont
  {Ruan}}]{Bao2012jqsrt}%
  \BibitemOpen
  \bibfield  {author} {\bibinfo {author} {\bibfnamefont {H.}~\bibnamefont
  {Bao}}, \bibinfo {author} {\bibfnamefont {B.}~\bibnamefont {Qiu}}, \bibinfo
  {author} {\bibfnamefont {Y.}~\bibnamefont {Zhang}}, \ and\ \bibinfo {author}
  {\bibfnamefont {X.}~\bibnamefont {Ruan}},\ }\href {\doibase
  10.1016/j.jqsrt.2012.04.018} {\bibfield  {journal} {\bibinfo  {journal}
  {Journal of Quantitative Spectroscopy and Radiative Transfer}\ }\textbf
  {\bibinfo {volume} {113}},\ \bibinfo {pages} {1683} (\bibinfo {year}
  {2012})}\BibitemShut {NoStop}%
\bibitem [{\citenamefont {Dove}(1993)}]{Dove_book}%
  \BibitemOpen
  \bibfield  {author} {\bibinfo {author} {\bibfnamefont {M.~T.}\ \bibnamefont
  {Dove}},\ }\href@noop {} {\emph {\bibinfo {title} {{Introduction to Lattice
  Dynamics}}}}\ (\bibinfo  {publisher} {Cambridge University Press},\ \bibinfo
  {address} {New York, USA},\ \bibinfo {year} {1993})\BibitemShut {NoStop}%
\bibitem [{\citenamefont {Qiu}\ and\ \citenamefont
  {Ruan}(2011)}]{Qiu2011arXiv}%
  \BibitemOpen
  \bibfield  {author} {\bibinfo {author} {\bibfnamefont {B.}~\bibnamefont
  {Qiu}}\ and\ \bibinfo {author} {\bibfnamefont {X.}~\bibnamefont {Ruan}},\
  }\href {http://arxiv.org/abs/1111.4613v1} {\bibfield  {journal} {\bibinfo
  {journal} {arXiv preprint arXiv:1111.4613}\ } (\bibinfo {year} {2011})},\
  \Eprint {http://arxiv.org/abs/arXiv:1111.4613v1} {arXiv:arXiv:1111.4613v1}
  \BibitemShut {NoStop}%
\bibitem [{\citenamefont {Feng}\ \emph {et~al.}(2015)\citenamefont {Feng},
  \citenamefont {Qiu},\ and\ \citenamefont {Ruan}}]{Feng2015jap}%
  \BibitemOpen
  \bibfield  {author} {\bibinfo {author} {\bibfnamefont {T.}~\bibnamefont
  {Feng}}, \bibinfo {author} {\bibfnamefont {B.}~\bibnamefont {Qiu}}, \ and\
  \bibinfo {author} {\bibfnamefont {X.}~\bibnamefont {Ruan}},\ }\href {\doibase
  10.1063/1.4921108} {\bibfield  {journal} {\bibinfo  {journal} {Journal of
  Applied Physics}\ }\textbf {\bibinfo {volume} {117}},\ \bibinfo {pages}
  {195102} (\bibinfo {year} {2015})}\BibitemShut {NoStop}%
\bibitem [{\citenamefont {de~Koker}(2009)}]{Koker2009prl}%
  \BibitemOpen
  \bibfield  {author} {\bibinfo {author} {\bibfnamefont {N.}~\bibnamefont
  {de~Koker}},\ }\href {\doibase 10.1103/PhysRevLett.103.125902} {\bibfield
  {journal} {\bibinfo  {journal} {Physical Review Letters}\ }\textbf {\bibinfo
  {volume} {103}},\ \bibinfo {pages} {125902} (\bibinfo {year}
  {2009})}\BibitemShut {NoStop}%
\bibitem [{\citenamefont {Thomas}\ \emph {et~al.}(2010)\citenamefont {Thomas},
  \citenamefont {Turney}, \citenamefont {Iutzi}, \citenamefont {Amon},\ and\
  \citenamefont {McGaughey}}]{Thomas2010prb}%
  \BibitemOpen
  \bibfield  {author} {\bibinfo {author} {\bibfnamefont {J.~A.}\ \bibnamefont
  {Thomas}}, \bibinfo {author} {\bibfnamefont {J.~E.}\ \bibnamefont {Turney}},
  \bibinfo {author} {\bibfnamefont {R.~M.}\ \bibnamefont {Iutzi}}, \bibinfo
  {author} {\bibfnamefont {C.~H.}\ \bibnamefont {Amon}}, \ and\ \bibinfo
  {author} {\bibfnamefont {A.~J.~H.}\ \bibnamefont {McGaughey}},\ }\href
  {\doibase 10.1103/PhysRevB.81.081411} {\bibfield  {journal} {\bibinfo
  {journal} {Physical Review B}\ }\textbf {\bibinfo {volume} {81}},\ \bibinfo
  {pages} {081411} (\bibinfo {year} {2010})}\BibitemShut {NoStop}%
\bibitem [{\citenamefont {{Zwanzig}}(1965)}]{GreenKubo}%
  \BibitemOpen
  \bibfield  {author} {\bibinfo {author} {\bibfnamefont {R.}~\bibnamefont
  {{Zwanzig}}},\ }\href {\doibase 10.1146/annurev.pc.16.100165.000435}
  {\bibfield  {journal} {\bibinfo  {journal} {Annual Review of Physical
  Chemistry}\ }\textbf {\bibinfo {volume} {16}},\ \bibinfo {pages} {67}
  (\bibinfo {year} {1965})}\BibitemShut {NoStop}%
\end{thebibliography}%

\end{document}